\documentclass[aps,twocolumn,prb,preprintnumbers,amsmath,amssymb]{revtex4}
\usepackage{graphicx,amsmath,amsfonts,amssymb,color,ulem,bm,tabularx,sidecap,dcolumn,multirow}
\usepackage{times}
\usepackage[sort&compress]{natbib}
\usepackage[T1]{fontenc}

\usepackage{graphicx}
\usepackage{dcolumn}
\usepackage{slashbox}
\usepackage{bm}
\usepackage{color}
\usepackage{ulem}

\usepackage{braket}

\begin{document}

\title{Microscopic Description of Electric and Magnetic Toroidal Multipoles in Hybrid Orbitals}

\author{Satoru Hayami$^1$ and Hiroaki Kusunose$^2$}
 \affiliation{$^1$Department of Physics, Hokkaido University, Sapporo 060-0810, Japan \\
 $^2$Department of Physics, Meiji University, Kawasaki 214-8571, Japan}

\begin{abstract}
We present a general formalism of multipole descriptions under the space-time inversion group.
We elucidate that two types of atomic toroidal multipoles, i.e., electric and magnetic, are fundamental pieces to express electronic order parameters in addition to ordinary electric and magnetic multipoles.
By deriving quantum-mechanical operators for both toroidal multipoles, we show that electric (magnetic) toroidal multipole higher than dipole (monopole) can become a primary order parameter in a hybridized-orbital system. 
We also demonstrate emergent cross-correlated couplings between electric, magnetic, and elastic degrees of freedom, such as magneto-electric and magneto(electro)-elastic couplings, under toroidal multipole orders.  
\end{abstract}

\maketitle

Mutual interplay between fundamental degrees of freedom of electrons in solids, i.e., charge, spin, and orbital, has attracted growing interest in various context. 
The concept of atomic-scale multipole has been developed to describe such complex electronic degrees of freedom in a unified manner~\cite{Santini_RevModPhys.81.807,kuramoto2009multipole}. 
Especially, in $f$-electron systems, higher-rank multipoles have been introduced to describe peculiar ordered phases, e.g., electric quadrupole in Pr$T_2X_{20}$($T=$Ir, Rh, V, Ti and $X=$Al, Zn)~\cite{onimaru2016exotic}, magnetic octupole in Ce$_{1-x}$La$_x$B$_6$~\cite{Mannix_PhysRevLett.95.117206}, and so on. 
Meanwhile, in $d$-electron systems, an atomic multipole has been extended to an object that is defined over a cluster consisting of several atomic sites; magnetic monopole excitations in spin ice~\cite{castelnovo2008magnetic,khomskii2012electric}, magnetic octupole by noncollinear/noncoplanar magnetic structures~\cite{Arima_doi:10.7566/JPSJ.82.013705,Suzuki_PhysRevB.95.094406}, nematic (quadrupole) order in iron-based superconductors~\cite{chuang2010nematic,fernandes2014drives}, and spin chirality accompanying Berry phase~\cite{taguchi2001spin}. 
These systematic description in terms of multipoles has provided a seamless and microscopic understanding of fundamental physical phenomena, such as the anomalous Hall effect~\cite{Loss_PhysRevB.45.13544,Ye_PhysRevLett.83.3737,Nagaosa_RevModPhys.82.1539} and magneto-electric responses~\cite{Fiebig0022-3727-38-8-R01,cheong2007multiferroics,KhomskiiPhysics.2.20}. 

Under the space-time inversion group, two-types of multipoles are often discussed, namely, electric multipole (E: polar/true tensor) and magnetic multipole (M: axial/pseudo tensor)~\cite{Kusunose_JPSJ.77.064710,Santini_RevModPhys.81.807,kuramoto2009multipole}. 
In addition to these ordinary multipoles, another two-types of multipoles are definable: one is the electric toroidal (ET) multipole and the other is the magnetic toroidal (MT) multipole, which have opposite parity from the ordinary ones under spatial inversion~\cite{dubovik1990toroid,Hlinka_PhysRevLett.113.165502,papasimakis2016electromagnetic,Hlinka_PhysRevLett.116.177602}.  
The most familiar toroidal multipole is the MT dipole. 
It was originally introduced as a configuration of static currents flowing on the surface of a torus~\cite{zel1958relation}, whose concept was developed into various fields, e.g., nuclear~\cite{Flambaum_PhysRevC.56.1641} and molecular physics~\cite{Ceulemans_PhysRevLett.80.1861}, and classical electrodynamics~\cite{dubovik1975multipole}. 
In condensed matter physics, the MT dipole has been extensively investigated due to its potential role for exotic phenomena, such as magneto-electric effect and nonreciprocal directional dichroism~\cite{gorbatsevich1994toroidal,popov1998magnetoelectric,Sawada_PhysRevLett.95.237402,EdererPhysRevB.76.214404,kopaev2009toroidal,Spaldin_0953-8984-20-43-434203,Hayami_PhysRevB.90.024432,hayami2016emergent,gao2017microscopic}. 
Such a MT dipole is often identified with a vortex-type magnetic orderings over {\it several atomic sites}. 
However, MT multipole should be evaluated even at {\it single atomic site} because it is independent of M multipole under the space-time inversion group.
It is then desirable to construct a microscopic description of toroidal multipole and clarify when it can be a primary order parameter characterizing thermodynamic phases in condensed matter.

In this Letter, we present a general formalism to describe not only MT but also ET multipoles from a single atomic viewpoint. 
By considering the correspondence between the classical description in the expansion of electromagnetic potentials and the quantum-mechanical operators, we obtain microscopic expressions of both ET and MT multipoles. 
We demonstrate that the atomic ET and MT multipoles can be activated in the Hilbert space spanned by orbitals with different azimuthal quantum number, e.g., $s$-$d$, $p$-$d$, and $d$-$f$ hybrid orbitals. 
The atomic ET and MT multipoles in addition to the ordinary E and M multipoles constitute a complete set to express an arbitrary degree of freedom in the hybrid orbitals. 
We also elucidate possible cross-correlated responses in the presence of an ET or MT multipole ordering.

Let us begin with the spatial distributions of the electric scalar potential $\phi(\bm{r})$ and magnetic vector potential $\bm{A}(\bm{r})$ in the presence of the electric charge density $\rho_{\rm e}(\bm{r})$ and electric current density $\bm{j}_{\rm e}(\bm{r})$. 
Under the Coulomb gauge $\bm{\nabla}\cdot \bm{A}(\bm{r}) =0$, it is well known that the static solutions of $\phi(\bm{r})$ and $\bm{A}(\bm{r})$ for the region outside the sources, $\rho_{\rm e}(\bm{r})$ and $\bm{j}_{\rm e}(\bm{r})$, are given in the form of the multipole expansion~\cite{Schwartz_PhysRev.97.380,dubovik1990toroid},
\begin{align}
\label{eq:phi}
&\phi (\bm{r}) = \sum_{lm} Q_{lm}\frac{C^Q_lY_{lm} (\hat{\bm{r}})}{r^{l+1}}  , \\
\label{eq:A}
&\bm{A} (\bm{r}) = \sum_{lm}\left[
 M_{lm}\frac{C^M_l\bm{Y}^l_{lm} (\hat{\bm{r}})}{r^{l+1}}+T_{lm}\frac{C^T_l\bm{Y}^{l+1}_{lm}  (\hat{\bm{r}})}{r^{l+2}}
\right], 
\end{align}
where the normalization constants are given by $C^Q_l=\sqrt{4 \pi/( 2 l +1)}$, $C^M_l=i\sqrt{4 \pi(l+1)/( 2 l +1)l}$, and $C^T_l=-\sqrt{4\pi(l+1)}$, respectively. 
$Y_{lm}(\hat{\bm{r}})$ [$\bm{Y}^{l'}_{lm}(\hat{\bm{r}})$] as a function of angles $\hat{\bm{r}}=\bm{r}/r$ denotes the spherical (vector spherical) harmonics~\cite{Blatt1991,Kusunose_JPSJ.77.064710}, where $l$ and $m$ are the azimuthal and magnetic quantum numbers, respectively ($-l\leq m \leq l$ and $l'=l, l \pm 1$). 
Note that $l'=l-1$ component does not appear in Eq.~(\ref{eq:A}) due to the Coulomb gauge. 
$Q_{lm}$ in Eq.~(\ref{eq:phi}) represents E multipole that is given by 
\begin{align}
\label{eq:Emultipole}
&Q_{lm}= \int d\bm{r} \rho_{\rm e}(\bm{r}) O_{lm}(\bm{r}), 
\end{align}
characterizing anisotropy of electric distribution.
We introduced $O_{lm} (\bm{r})=  C^Q_l r^l Y^*_{lm}(\bm{\hat{r}})$ for notational simplicity. 
Similarly, $M_{lm}$ and $T_{lm}$ in Eq.~(\ref{eq:A}) are M and MT multipoles, 
\begin{align}
\label{eq:Mmultipole}
&M_{lm}= \frac{1}{c(l+1)} \int d\bm{r} [\bm{r} \times \bm{j}_{\rm e}(\bm{r})] \cdot \bm{\nabla}O_{lm}(\bm{r}), \\
\label{eq:MTmultipole}
&T_{lm}= \frac{1}{c(l+1)} \int d\bm{r} [\bm{r}\cdot \bm{j}_{\rm e}(\bm{r})] O_{lm}(\bm{r}), 
\end{align}
respectively.
As $O_{lm}(\bm{r})$ has the parity $(-1)^l$ under spatial inversion, $Q_{lm}$ and $T_{lm}$ have the parity $(-1)^{l}$ (polar tensor), while $M_{lm}$ has the parity $(-1)^{l+1}$ (axial tensor). 
It should be noted that the multipole expansion does not contain the ET multipoles as well as the M and MT monopoles, $M_{00}$ and $T_{00}$, since there are no magnetic charge $\rho_{\rm m}(\bm{r})$ and magnetic (spin) current $\bm{j}_{\rm m} (\bm{r})$ as fundamental sources. 

These multipoles are also expressed by using polarization densities: electric polarization $\bm{P}(\bm{r})$, magnetization $\bm{M}(\bm{r})$, and magnetic toroidalization $\bm{T}(\bm{\bm{r}})$. 
They are related with $\rho_{\rm e}(\bm{r})$ and $\bm{j}_{\rm e}(\bm{r})$ as follows: 
$\rho_{\rm e} (\bm{r}) = -\bm{\nabla}\cdot \bm{P}(\bm{r})$, 
$\bm{j}_{\rm e} (\bm{r}) = c [\bm{\nabla}\times \bm{M}(\bm{r})]$, and 
$\bm{M} (\bm{r}) = \bm{\nabla}\times \bm{T}(\bm{r})$.  
Then, Eqs.~(\ref{eq:Emultipole})-(\ref{eq:MTmultipole}) are rewritten as 
\begin{align}
\label{eq:Emultipole2}
&Q_{lm} = \int d\bm{r} \ \bm{P}(\bm{r}) \cdot \bm{\nabla}O_{lm}(\bm{r}), \\
\label{eq:Mmultipole2}
&M_{lm} = \int d\bm{r} \ \bm{M}(\bm{r}) \cdot \bm{\nabla}O_{lm}(\bm{r}), \\
\label{eq:MTmultipole2}
&T_{lm}= \int d\bm{r} \ \bm{T}(\bm{r}) \cdot \bm{\nabla}O_{lm}(\bm{r}), 
\end{align}
by means of the identity for arbitrary vector field $\bm{X}(\bm{r})$,
\begin{align}
\label{eq:identity}
\int d\bm{r}\,\frac{\bm{r}\times (\bm{\nabla}\times \bm{X})}{l+1}\cdot\bm{\nabla} O_{lm}
= \int d\bm{r}\,\bm{X} \cdot \bm{\nabla} O_{lm}. 
\end{align}

We now turn to a description of ET multipoles. 
As was mentioned, ET multipoles do not appear in the multipole expansion of electromagnetic potentials. 
Nevertheless, ET multipoles should exist under the space-time inversion group, as the multipole degree of freedom corresponding to time-reversal-even axial tensor is missing. 

In order to obtain an expression of ET multipoles, we focus on a dual nature between electric and magnetic degrees of freedom in the presence of magnetic current density $\bm{j}_{\rm m} (\bm{r})$.
Although $\bm{j}_{\rm m}(\bm{r})$ does not exist in fundamental level, a vorticity of the electric polarization $\bm{P}(\bm{r})$ corresponds to $\bm{j}_{\rm m} (\bm{r})$ as similar to the electric current density as a vorticity of $\bm{M}(\bm{r})$.
Indeed, the magnetic current density is defined by $\bm{j}_{\rm m} (\bm{r}) = c [\bm{\nabla}\times \bm{P}(\bm{r})]$.
The transformation of $\bm{j}_{\rm e} (\bm{r}) \to \bm{j}_{\rm m} (\bm{r})$ reverses both the time-reversal and spatial inversion properties. 
Thus, by replacing $\bm{j}_{\rm e} (\bm{r})$ with $\bm{j}_{\rm m} (\bm{r})$ in the expression of $M_{lm}$, Eq.~(\ref{eq:Mmultipole}), and using the identity Eq.~(\ref{eq:identity}), we again obtain the expression of $Q_{lm}$ in Eq.~(\ref{eq:Emultipole2}) as it is. 

The above consideration implies that the ET multipoles can be obtained from the MT multipoles by the transformation of $\bm{j}_{\rm e} (\bm{r}) \rightarrow \bm{j}_{\rm m} (\bm{r})$. 
With use of Eqs.~(\ref{eq:MTmultipole}) and (\ref{eq:identity}), the ET multipole is given by 
\begin{align}
G_{lm} &= \frac{1}{c(l+1)} \int d \bm{r}  [\bm{r} \cdot \bm{j}_{\rm m}(\bm{r})] O_{lm}(\bm{r}) \cr
            &= \frac{1}{l+1} \int d\bm{r}  [\bm{r}\times \bm{P}(\bm{r})] \cdot \bm{\nabla} O_{lm}(\bm{r}) \cr
\label{eq:ETmultipole}
            &= \int d\bm{r} \ \bm{G}(\bm{r})\cdot \bm{\nabla}O_{lm}(\bm{r}),
\end{align}
where $\bm{P} (\bm{r}) = \bm{\nabla}\times \bm{G}(\bm{r})$ and $\bm{G}(\bm{r})$ is the electric toroidalization. 
The expression of $G_{lm}$ in Eq.~(\ref{eq:ETmultipole}) clearly exhibits a time-reversal-even axial tensor. 
Moreover, the form of $\bm{r}\times \bm{P}(\bm{r})$ in the second line of Eq.~(\ref{eq:ETmultipole}) is consistent with the phenomenological description of the ET dipole as a toroidal alignment of electric dipoles~\cite{naumov2004unusual,Prosandeev_PhysRevLett.96.237601}.  
We summarize four fundamental multipoles in Table~\ref{tab1}.

\begin{table}[t!]
\begin{center}
\caption{
Four fundamental multipoles under the space-time inversion group. 
$\mathcal{T}$ and $\mathcal{P}$ represent the time-reversal and spatial inversion operations. 
}

\label{tab1}
\begingroup
\renewcommand{\arraystretch}{1.2}
 \begin{tabular}{lcclcccccc}
 \hline \hline
type & notation & $\mathcal{T}$ & $\mathcal{P}$ 
& source  & pol. & remark   
 \\
\hline
E  & $Q_{lm}$ & $+$ & $(-1)^l$     & $\rho_{\rm e}$, $\bm{j}_{\rm m}$ & $\bm{P}$ & $\rho_{\rm e} = - \bm{\nabla \cdot \bm{P}}$\\ 
M  & $M_{lm}$ & $-$ & $(-1)^{l+1}$ & $\bm{j}_{\rm e}$ & $\bm{M}$ & $\bm{j}_{\rm e}=c (\bm{\nabla}\times \bm{M})$ \\ 
MT & $T_{lm}$ & $-$ & $(-1)^l$     & $\bm{j}_{\rm e}$  & $\bm{T}$ & $\bm{M}= \bm{\nabla}\times \bm{T}$     \\ 
ET & $G_{lm}$ & $+$ & $(-1)^{l+1}$ & $\bm{j}_{\rm m}$  & $\bm{G}$ &$\bm{P}= \bm{\nabla}\times \bm{G}$ \\ \hline\hline
\end{tabular}
\endgroup
\end{center}
\end{table}

\begin{table*}[htb!]
\begin{center}
\caption{
Operator expressions of multipoles up to $l=3$. 
$-e$ ($-\mu_{\rm B}$) is the unit for E/ET (M/MT) multipoles, which are classified by the irreducible representations of the cubic $O$ group.
For a noncommute product, it should be regarded as a symmetrized expression, e.g., $AB\to (AB+B^{\dagger}A^{\dagger})/2$. 
The M multipole operators (e.g. $M_{xyz}$) are obtained by replacing $\bm{t}_l$ in the MT ones (e.g. $T_{xyz}$) with $\bm{m}_l$.
The ``elementary'' sources, $\bm{m}_{l}$, $\bm{t}_{l}$, and $g^{\alpha\beta}_{l}$ are defined in the main text. 
}
\label{tab2}
\begingroup
\renewcommand{\arraystretch}{1.2}
 \begin{tabular}{lccccccccc}
 \hline \hline
$l$ &type & $\mathcal{T}$ & $\mathcal{P}$ & irrep. & symbol & definition \\ \hline
$0$ & E  & $+$ & $+$ & $\Gamma_1$ $(A_1)$ & $Q_0$ & $1$ \\  \hline
$1$ & E  & $+$ & $-$ & $\Gamma_4$ $(T_1)$ & $Q_x$, $Q_y$, $Q_z$ & $x$, $y$, $z$ \\
    & MT & $-$ & $-$ & $\Gamma_4$ $(T_1)$ & $T_x$, $T_y$, $T_z$ & $t_{1}^{x}$, $t_{1}^{y}$, $t_{1}^{z}$ \\ \hline
$2$ & E  & $+$ & $+$ & $\Gamma_3$ $(E)$ & $Q_{u}$, $Q_{v}$ & $\frac{1}{2}(3z^2-r^2)$, $\frac{\sqrt{3}}{2}(x^2-y^2)$ \\
    &    &     &     & $\Gamma_5$ $(T_2)$ & $Q_{yz}$, $Q_{zx}$, $Q_{xy}$ & $\sqrt{3}yz$, $\sqrt{3}zx$, $\sqrt{3}xy$ \\
    & MT & $-$ & $+$ & $\Gamma_3$ $(E)$ & $T_{u}$, $T_{v}$ & $3zt_{2}^{z}-\bm{r}\cdot \bm{t}_2$, $\sqrt{3} (xt_{2}^{x}-yt_{2}^{y})$ \\
    &    &     &     & $\Gamma_5$ $(T_2)$ & $T_{yz}$, $T_{zx}$, $T_{xy}$ & $\sqrt{3}(yt_{2}^{z}+zt_{2}^{y})$, $\sqrt{3}(zt_{2}^{x}+xt_{2}^{z})$, $\sqrt{3}(xt_{2}^{y}+yt_{2}^{x})$ \\ 
    & ET & $+$ & $-$ & $\Gamma_3$ $(E)$ & $G_{u}$, $G_{v}$ & $3g_{2}^{zz} -\sum_{\alpha}g_{2}^{\alpha\alpha}$, $\sqrt{3}(g_{2}^{xx}-g_{2}^{yy})$ \\
    &    &     &     & $\Gamma_5$ $(T_2)$ & $G_{yz}$, $G_{zx}$, $G_{xy}$ & $2\sqrt{3}g_{2}^{yz}$, $2\sqrt{3}g_{2}^{zx}$, $2\sqrt{3}g_{2}^{xy}$ \\ \hline
3 & E & $+$ & $-$ & $\Gamma_{2}$ $(A_{2})$ & $Q_{xyz}$ & $\sqrt{15}xyz$ \\
  &   & & & $\Gamma_{4}$ ($T_{1}$) & $Q_{x}^{\alpha}$, $Q_{y}^{\alpha}$, $Q_{z}^{\alpha}$ & $\frac{1}{2}x(5x^{2}-3r^{2})$, (cyclic) \\
  &   & & & $\Gamma_{5}$ ($T_{2}$) & $Q_{x}^{\beta}$, $Q_{y}^{\beta}$, $Q_{z}^{\beta}$ & $\frac{\sqrt{15}}{2}x(y^{2}-z^{2})$, (cyclic) \\
  & MT & $-$ & $-$ & $\Gamma_{2}$ ($A_{2}$) & $T_{xyz}$ & $\sqrt{15}(yzt_{3}^{x}+zxt_{3}^{y}+xyt_{3}^{z})$ \\
  &   & & & $\Gamma_{4}$ ($T_{1}$) & $T_{x}^{\alpha}$, $T_{y}^{\alpha}$, $T_{z}^{\alpha}$ & $3[\frac{1}{2}(3x^{2}-r^{2})t_{3}^{x}-x(yt_{3}^{y}+zt_{3}^{z})]$, (cyclic) \\
  &   & & & $\Gamma_{5}$ ($T_{2}$) & $T_{x}^{\beta}$, $T_{y}^{\beta}$, $T_{z}^{\beta}$ & $\sqrt{15}[\frac{1}{2}(y^{2}-z^{2})t_{3}^{x}+x(yt_{3}^{y}-zt_{3}^{z})]$, (cyclic) \\
  & ET & $+$ & $+$ & $\Gamma_{2}$ ($A_{2}$) & $G_{xyz}$ & $2\sqrt{15}(xg_{3}^{yz}+yg_{3}^{zx}+zg_{3}^{xy})$ \\
  &   & & & $\Gamma_{4}$ ($T_{1}$) & $G_{x}^{\alpha}$, $G_{y}^{\alpha}$, $G_{z}^{\alpha}$ & $9xg_{3}^{xx}-6(yg_{3}^{xy}+zg_{3}^{zx})-3x\sum_{\alpha}g_{3}^{\alpha\alpha}$, (cyclic) \\
  &   & & & $\Gamma_{5}$ ($T_{2}$) & $G_{x}^{\beta}$, $G_{y}^{\beta}$, $G_{z}^{\beta}$ & $\sqrt{15}[2(yg_{3}^{xy}-zg_{3}^{zx})+x(g_{3}^{yy}-g_{3}^{zz})]$, (cyclic) \\ \hline\hline
\end{tabular}
\endgroup
\end{center}
\end{table*}

Next, let us derive a quantum-mechanical operator expression of each multipole. 
By substituting the one-body electric charge density operator, 
$\hat{\rho}_{\rm e} (\bm{r}) = -e\sum_j \delta (\bm{r}-\bm{r}_j)$, into Eq.~(\ref{eq:Emultipole}), the E multipole operator is obtained as 
\begin{align}
\label{eq:Emultipole_Q}
\hat{Q}_{lm} = -e\sum_j O_{lm} (\bm{r}_j),
\end{align}
where $\bm{r}_{j}$ is the position vector of each electron~\cite{Kusunose_JPSJ.77.064710,Santini_RevModPhys.81.807,kuramoto2009multipole}. 

In a similar manner, the M and MT multipole operators are obtained by using the electric current density operator $\hat{\bm{j}}_{\rm e}$ consisting of the orbital and spin parts, 
\begin{align}
\label{eq:currento_Q}
&\frac{1}{2c} (\bm{r}\times \hat{\bm{j}}^{\rm (o)}_{\rm e}) = - \mu_{\rm B} \sum_j \bm{l}_j  \delta (\bm{r}-\bm{r}_j), \\
\label{eq:currents_Q}
&\frac{1}{c} \hat{\bm{j}}^{\rm (s)}_{\rm e} = -\mu_{\rm B} \sum_{j} (\bm{\nabla}\times \bm{\sigma}_j)\delta (\bm{r}-\bm{r}_j), 
\end{align}
where $\mu_{\rm B}= e \hbar /2mc$ is the Bohr magneton, and $\bm{l}_j$ and $\bm{\sigma}_j$ are the orbital and spin angular-momentum operators of electron at $\bm{r}_{j}$. 
By substituting Eqs.~(\ref{eq:currento_Q}) and (\ref{eq:currents_Q}) into Eqs.~(\ref{eq:Mmultipole}) and (\ref{eq:MTmultipole}), and after some algebra, the M and MT multipole operators are obtained as~\cite{Kusunose_JPSJ.77.064710,Santini_RevModPhys.81.807,kuramoto2009multipole} 
\begin{align}
\label{eq:Mmultipole_Q}
\hat{M}_{lm} &= -\mu_{\rm B}\sum_j \bm{m}_{l}(\bm{r}_{j})\cdot\bm{\nabla} O_{lm}(\bm{r}_j),
\cr
&\bm{m}_{l}(\bm{r}_{j})=\frac{2 \bm{l}_j}{l+1} + \bm{\sigma}_j, \\
\label{eq:MTmultipole_Q}
\hat{T}_{lm} &=  -\mu_{\rm B}\sum_j \bm{t}_{l}(\bm{r}_{j})\cdot \bm{\nabla} O_{lm}(\bm{r}_j),
\cr
&\bm{t}_{l}(\bm{r}_{j})=\frac{\bm{r}_j}{l+1} \times \left( \frac{2\bm{l}_j}{l+2} + \bm{\sigma}_j \right).
\end{align}
Note that $\hat{M}_{lm}$ and $\hat{T}_{lm}$ vanishes for monopole, $l=0$, owing to the derivative of $O_{lm}$.

In order to examine the ET multipole operator, we again focus on the dual nature between electric and magnetic degrees of freedom.
To this end, let us consider a transformation that reverses the time-reversal property without changing its spatial parity.
This is done by the operator, $R_{T}\equiv \bm{t}_{l}\cdot\bm{\nabla}$.
For instance, by applying $R_{T}$ to $O_{lm}$, one of the position vector $\bm{r}$ in the polynomial $O_{lm}(\bm{r})$ is replaced with $\bm{t}_{l}$.
We indeed obtain $\hat{T}_{lm}$ in Eq.~(\ref{eq:MTmultipole_Q}) by applying $(\mu_{\rm B}/e)R_{T}$ to $\hat{Q}_{lm}$ in Eq.~(\ref{eq:Emultipole_Q}).

Similarly, by applying $(e/\mu_{\rm B})R_{T}$ to $\hat{M}_{lm}$, we obtain the operator expression of the ET multipole as
\begin{align}
\hat{G}_{lm}&=-e\sum_{j}\sum_{\alpha\beta}^{x,y,z}g_{l}^{\alpha\beta}\nabla_{\alpha}\nabla_{\beta} O_{lm}(\bm{r}_{j}),
\cr&
g_{l}^{\alpha\beta}(\bm{r}_{j})=m_{l}^{\alpha}(\bm{r}_{j})t_{l}^{\beta}(\bm{r}_{j}).
\label{eq:ETmultipole_Q}
\end{align}
$\hat{G}_{lm}$ vanishes for $l=0$, $1$ due to the second derivative of $O_{lm}$, namely, the lowest-rank atomic ET multipole is quadrupole.
Note that the normalization of $\hat{G}_{lm}$ is not uniquely determined in contrast to other three multipoles.
When we consider the contribution only from the orbital parts in $\bm{m}_{l}$ and $\bm{t}_{l}$ for simplicity, the ET multipole is rewritten as 
\begin{align}
\hat{G}_{lm} &= -e  \sum_j \bm{g}_{l}(\bm{r}_{j})\cdot\bm{\nabla} O_{lm}(\bm{r}_j),
\cr
&\bm{g}_{l}(\bm{r}_{j})=\frac{4i\,\bm{l}_j (\bm{l}_j \cdot \bm{l}_j)}{(l+1)^2 (l+2)}.
\end{align}
The specific expressions of four multipole operators up to $l=3$ are shown in Table~\ref{tab2}.

In order to clarify when the ET and MT multipoles at single atomic site are activated, we calculate a matrix element of the multipole operators in Eqs.~(\ref{eq:Emultipole_Q}), (\ref{eq:Mmultipole_Q}), (\ref{eq:MTmultipole_Q}), and (\ref{eq:ETmultipole_Q}). 
We here consider the most fundamental situation under the rotation group, where the basis wave functions are characterized by $s$, $p$, $d$, and $f$ orbitals with angular momentum $L=0$-$3$ with its magnetic quantum number $M$, and we omit the contribution from the spin parts.   
The following results are readily extended to the situations under any point group, since it is a subgroup of the rotation group. 
Moreover, taking account of the spin parts and use of total angular-momentum basis $(J,J_{z})$ are straightforward. 

Within the basis having definite angular momentum $L$ (non-hybrid orbitals), the ordinary E and M multipoles with even-parity and the rank less than $2L$ are active.
Thus, the even-rank E multipoles and odd-rank M multipoles are non-vanishing in this Hilbert space~\cite{Kusunose_JPSJ.77.064710,Santini_RevModPhys.81.807,kuramoto2009multipole}. 

\begin{table}[t!]
\begin{center}
\caption{
Active multipoles in non-hybrid (intra) and hybrid (inter) orbitals. 
The number of independent multipoles to span the relevant Hilbert space is indicated in the parenthesis.
$\mathcal{P}$ represents the parity of the active multipoles with the rank $l$.
}
\label{tab3}
\begingroup
\renewcommand{\arraystretch}{1.2}
 \begin{tabular}{lccccccccc}
 \hline \hline
 \multicolumn{2}{l}{basis} &$\mathcal{P}$ & $l=0$ & $l=1$ & $l=2$ & $l=3$ & $l=4$ & $l=5$ & $l=6$     \\
\hline
$s$-$s$ &(1)& $+$    &E& --& --&--&-- &--&--\\ 
$p$-$p$ &(9)&           &E&M&E&--&--&--&--\\ 
$d$-$d$ &(25)&         &E&M&E&M&E&--&--      \\ 
$f$-$f$ &(49)&           &E&M&E&M&E&M&E \\ \hline
$s$-$d$ &(10)& $+$   &--&--&E/MT&--&--&--&-- \\ 
$p$-$f$ &(42)&           &--&--&E/MT&M/ET&E/MT&--&--\\ \hline
$s$-$p$ &(6)& $-$      &--&E/MT&--&--&--&--&-- \\ 
$s$-$f$ &(14)&           & --&--&--&E/MT&--&-- &--\\ 
$p$-$d$ &(30)&          &--&E/MT&M/ET&E/MT &--&--&--\\ 
$d$-$f$ &(70)&           & --&E/MT&M/ET&E/MT&M/ET&E/MT&--      \\ \hline\hline
\end{tabular}
\endgroup
\end{center}
\end{table}

On the other hand, the ET and MT multipoles are activated in hybrid orbitals. 
In the even-parity hybridization like $s$-$d$ and $p$-$f$ orbitals, the odd-rank ET multipoles and even-rank MT multipoles are activated, while in the odd-parity hybridization like $s$-$p$, $s$-$f$, $p$-$d$, and $d$-$f$ orbitals, the even-rank ET multipoles and the odd-rank MT multipoles are activated.
In addition to these toroidal multipoles, the ordinary E and M multipoles are also active, and the E and MT or M and ET multipoles appear as a pair.
It is noted that the matrix elements of a pair of multipoles are proportional with each other, namely, 
\begin{align}
&\langle LM | \hat{T}_{lm} | L'M' \rangle \propto \pm i \langle LM | \hat{Q}_{lm} | L'M' \rangle \ \  (L \lessgtr L'), \\
&\langle LM | \hat{G}_{lm} | L'M' \rangle \propto \pm  i \langle LM | \hat{M}_{lm} | L'M' \rangle \ \  (L \lessgtr L').
\end{align}
This relation is ascribed to the fact that the E (M) and MT (ET) multipoles have the same spatial inversion property, and differ only the time-reversal property~\cite{toroidal1}.  
All these multipoles constitute a complete set in the relevant Hilbert space, since the total number of independent multipoles is equal to that of the independent matrix elements.
We summarize all the active multipoles in the hybrid orbitals as well as the non-hybrid orbitals in Table~\ref{tab3}~\cite{toroidal1}.

\begin{table}[t!]
\begin{center}
\caption{
Cross correlations under a toroidal multipole ordering.
The induced multipoles by applying electric ($\bm{E}$), magnetic ($\bm{B}$), and rank-2 strain ($\varepsilon_{\alpha\beta}$) fields are listed. 
The number in the bracket represents the rank of corresponding multipoles.
}
\label{tab4}
\begingroup
\renewcommand{\arraystretch}{1.2}
 \begin{tabular}{cccccc}
 \hline \hline
order & basis & $\mathcal{P}$  & $\bm{E}$ & $\bm{B}$ & $\varepsilon_{\alpha\beta}$ \\
\hline
MT[2] & $s$-$d$ & $+$ & --- & E[2] & M[1,3], MT[2]
   \\ 
 & $p$-$f$ &  & --- & E[2], ET[3] & M[1,3], MT[2]
   \\ 
ET[3] & $p$-$f$ & & --- & M[3], MT[2] & E[2], ET[3]
   \\ 
\hline
MT[1] & $s$-$p$ & $-$& M[1]   & E[1]& MT[1]
   \\ 
 & $p$-$d$, $d$-$f$ & & M[1,3]   & E[1,3], ET[2]& MT[1,3], M[2]
   \\ 
MT[3] & $s$-$f$ & & ---  & E[3]& MT[3]
   \\ 
 & $p$-$d$, $d$-$f$ & & M[1,3] & E[1,3], ET[2]& MT[1,3], M[2]
   \\ 
ET[2] & $p$-$d$, $d$-$f$ & & E[2]   & MT[1,3], M[2]& E[1,3], ET[2]
   \\ 
\hline
\hline
\end{tabular}
\endgroup
\end{center}
\end{table}

Finally, we discuss possible cross-correlated couplings in applied external fields under an ET or MT multipole ordering. 
The most familiar response is the magneto-electric (ME) coupling where the magnetization (electric polarization) is induced by external electric $\bm{E}$ (magnetic $\bm{B}$) field. 
An atomic MT multipole order indeed gives rise to the ME coupling. 
For example, $M_y$ ($Q_y$) is induced by $E_z$ ($B_z$) under the MT dipole $T_x$ order in the odd-parity hybrid orbitals. 
Another example is the magneto(electro)-elastic coupling where the magnetization (electric polarization) is induced by external strain field $\varepsilon_{\alpha\beta}$ using ultrasonic wave and vice versa. 
MT (ET) quadrupole orderings exhibit such a cross correlation, for example, $M_x$ ($Q_x$) is induced by the $Q_{u}$-type strain field under the $T_{yz}$ ($G_{yz}$) toroidal quadrupole order. 
It is noteworthy to point out that there is a variety of cross-correlated couplings with underlying MT and/or ET multipole orders, e.g., the MT dipole $T_z$ is induced by $B_y$ field in the $G_{yz}$ order, and so on. 
We summarize what types of cross correlations occur under electric, magnetic, and rank-2 strain fields in each hybrid orbitals in Table~\ref{tab4}~\cite{toroidal3}.

In summary, we have formulated the microscopic description of electric and magnetic toroidal multipoles under the space-time inversion group. 
Our study underscores that the toroidal multipoles can be activated even at single atomic site in a hybridized-orbital space, which may bring various cross-correlated couplings, such as magneto-electric and magneto(electro)-elastic couplings. 
The concept of toroidal multipoles will be widely utilized at different length scales, since hybrid orbitals are ubiquitously inherent in various fields of physics, such as nuclear, molecular, nano, and solid-state physics, which are irrespective of metallic and insulating systems. 
The potential realizations of the microscopic toroidal multipoles are (1) strongly hybridized $f$-electron systems like U-based or cage compounds~\cite{tanida2010possible,mydosh2014hidden,onimaru2016exotic}, (2) topological semiconductors and excitonic insulators with different orbital characters of valence/conduction bands~\cite{Tsubouchi_PhysRevB.66.052418,Hasan_RevModPhys.82.3045,Kunes_PhysRevB.90.235112,Kaneko_PhysRevB.94.125127,yamaguchi2017multipole}, and (3) cluster systems including quantum dots and organic molecules~\cite{Hanson_RevModPhys.79.1217,Greene_RevModPhys.89.035006}.

We thank Y. Motome, J. Nasu, Y. Yanagi, T. Yanagisawa, and T. Kurumaji for fruitful discussions. 
This research was supported by JSPJ KAKENHI Grants Numbers 15K05176, 15H05885 (J-Physics), and 16H06590. 


\let\oldaddcontentsline\addcontentsline
\renewcommand{\addcontentsline}[3]{}

\let\addcontentsline\oldaddcontentsline

\newpage

\begin{widetext}
\appendix
\tableofcontents

\newpage
\section{Table of multipole operators}

\begin{table*}[htb!]
\begin{center}
\caption{
Operator expressions of multipoles up to $l=3$. 
$-e$ ($-\mu_{\rm B}$) is the unit for E/ET (M/MT) multipoles, which are classified by the irreducible representations of the cubic $O$ group.
For a noncommute product, it should be regarded as a symmetrized expression, e.g., $AB\to (AB+B^{\dagger}A^{\dagger})/2$.
The ``elementary'' sources, $\bm{m}_{l}$, $\bm{t}_{l}$, and $g^{\alpha\beta}_{l}$ are defined in the manuscript. 
}
\label{tab7}
\begingroup
\renewcommand{\arraystretch}{2.0}
 \begin{tabular}{lccccccccc}
 \hline \hline
$l$ &type & $\mathcal{T}$ & $\mathcal{P}$ & irrep. & symbol & definition \\ \hline
$0$ & E  & $+$ & $+$ & $\Gamma_1$ $(A_1)$ & $Q_0$ & $1$ \\  \hline
$1$ & E  & $+$ & $-$ & $\Gamma_4$ $(T_1)$ & $Q_x$, $Q_y$, $Q_z$ & $x$, $y$, $z$ \\
    & M  & $-$ & $+$ & $\Gamma_4$ $(T_1)$ & $M_x$, $M_y$, $M_z$ & $m_{1}^{x}$, $m_{1}^{y}$, $m_{1}^{z}$ \\
    & MT & $-$ & $-$ & $\Gamma_4$ $(T_1)$ & $T_x$, $T_y$, $T_z$ & $t_{1}^{x}$, $t_{1}^{y}$, $t_{1}^{z}$ \\ \hline
$2$ & E  & $+$ & $+$ & $\Gamma_3$ $(E)$ & $Q_{u}$, $Q_{v}$ & $\frac{1}{2}(3z^2-r^2)$, $\frac{\sqrt{3}}{2}(x^2-y^2)$ \\
    &    &     &     & $\Gamma_5$ $(T_2)$ & $Q_{yz}$, $Q_{zx}$, $Q_{xy}$ & $\sqrt{3}yz$, $\sqrt{3}zx$, $\sqrt{3}xy$ \\
    & M  & $-$ & $-$ & $\Gamma_3$ $(E)$ & $M_{u}$, $M_{v}$ & $3zm_{2}^{z}-\bm{r}\cdot \bm{m}_2$, $\sqrt{3} (xm_{2}^{x}-ym_{2}^{y})$ \\
    &    &     &     & $\Gamma_5$ $(T_2)$ & $M_{yz}$, $M_{zx}$, $M_{xy}$ & $\sqrt{3}(ym_{2}^{z}+zm_{2}^{y})$, $\sqrt{3}(zm_{2}^{x}+xm_{2}^{z})$, $\sqrt{3}(xm_{2}^{y}+ym_{2}^{x})$ \\
    & MT & $-$ & $+$ & $\Gamma_3$ $(E)$ & $T_{u}$, $T_{v}$ & $3zt_{2}^{z}-\bm{r}\cdot \bm{t}_2$, $\sqrt{3} (xt_{2}^{x}-yt_{2}^{y})$ \\
    &    &     &     & $\Gamma_5$ $(T_2)$ & $T_{yz}$, $T_{zx}$, $T_{xy}$ & $\sqrt{3}(yt_{2}^{z}+zt_{2}^{y})$, $\sqrt{3}(zt_{2}^{x}+xt_{2}^{z})$, $\sqrt{3}(xt_{2}^{y}+yt_{2}^{x})$ \\ 
    & ET & $+$ & $-$ & $\Gamma_3$ $(E)$ & $G_{u}$, $G_{v}$ & $3g_{2}^{zz} -\sum_{\alpha}g_{2}^{\alpha\alpha}$, $\sqrt{3}(g_{2}^{xx}-g_{2}^{yy})$ \\
    &    &     &     & $\Gamma_5$ $(T_2)$ & $G_{yz}$, $G_{zx}$, $G_{xy}$ & $2\sqrt{3}g_{2}^{yz}$, $2\sqrt{3}g_{2}^{zx}$, $2\sqrt{3}g_{2}^{xy}$ \\ \hline
3 & E & $+$ & $-$ & $\Gamma_{2}$ $(A_{2})$ & $Q_{xyz}$ & $\sqrt{15}xyz$ \\
  &   & & & $\Gamma_{4}$ ($T_{1}$) & $Q_{x}^{\alpha}$, $Q_{y}^{\alpha}$, $Q_{z}^{\alpha}$ & $\frac{1}{2}x(5x^{2}-3r^{2})$, (cyclic) \\
  &   & & & $\Gamma_{5}$ ($T_{2}$) & $Q_{x}^{\beta}$, $Q_{y}^{\beta}$, $Q_{z}^{\beta}$ & $\frac{\sqrt{15}}{2}x(y^{2}-z^{2})$, (cyclic) \\
  & M & $-$ & $+$ & $\Gamma_{2}$ ($A_{2}$) & $M_{xyz}$ & $\sqrt{15}(yzm_{3}^{x}+zxm_{3}^{y}+xym_{3}^{z})$ \\
  &   & & & $\Gamma_{4}$ ($T_{1}$) & $M_{x}^{\alpha}$, $M_{y}^{\alpha}$, $M_{z}^{\alpha}$ & $3[\frac{1}{2}(3x^{2}-r^{2})m_{3}^{x}-x(ym_{3}^{y}+zm_{3}^{z})]$, (cyclic) \\
  &   & & & $\Gamma_{5}$ ($T_{2}$) & $M_{x}^{\beta}$, $M_{y}^{\beta}$, $M_{z}^{\beta}$ & $\sqrt{15}[\frac{1}{2}(y^{2}-z^{2})m_{3}^{x}+x(ym_{3}^{y}-zm_{3}^{z})]$, (cyclic) \\
  & MT & $-$ & $-$ & $\Gamma_{2}$ ($A_{2}$) & $T_{xyz}$ & $\sqrt{15}(yzt_{3}^{x}+zxt_{3}^{y}+xyt_{3}^{z})$ \\
  &   & & & $\Gamma_{4}$ ($T_{1}$) & $T_{x}^{\alpha}$, $T_{y}^{\alpha}$, $T_{z}^{\alpha}$ & $3[\frac{1}{2}(3x^{2}-r^{2})t_{3}^{x}-x(yt_{3}^{y}+zt_{3}^{z})]$, (cyclic) \\
  &   & & & $\Gamma_{5}$ ($T_{2}$) & $T_{x}^{\beta}$, $T_{y}^{\beta}$, $T_{z}^{\beta}$ & $\sqrt{15}[\frac{1}{2}(y^{2}-z^{2})t_{3}^{x}+x(yt_{3}^{y}-zt_{3}^{z})]$, (cyclic) \\
  & ET & $+$ & $+$ & $\Gamma_{2}$ ($A_{2}$) & $G_{xyz}$ & $2\sqrt{15}(xg_{3}^{yz}+yg_{3}^{zx}+zg_{3}^{xy})$ \\
  &   & & & $\Gamma_{4}$ ($T_{1}$) & $G_{x}^{\alpha}$, $G_{y}^{\alpha}$, $G_{z}^{\alpha}$ & $9xg_{3}^{xx}-6(yg_{3}^{xy}+zg_{3}^{zx})-3x\sum_{\alpha}g_{3}^{\alpha\alpha}$, (cyclic) \\
  &   & & & $\Gamma_{5}$ ($T_{2}$) & $G_{x}^{\beta}$, $G_{y}^{\beta}$, $G_{z}^{\beta}$ & $\sqrt{15}[2(yg_{3}^{xy}-zg_{3}^{zx})+x(g_{3}^{yy}-g_{3}^{zz})]$, (cyclic) \\ \hline\hline
\end{tabular}
\endgroup
\end{center}
\end{table*}

\newpage
\section{Table of active multipoles}
We summarize activated multipoles in both non-hybrid and hybrid orbitals. 

\begin{align}
\begin{array}{l|l|l|l|l}
\text{basis}&s & p & d & f \\ \hline
s& Q_0 & Q_x, Q_y, Q_z & Q_{u},Q_{v},Q_{yz},Q_{zx},Q_{xy}&Q_{xyz}, Q_{x}^\alpha, Q_{y}^\alpha, Q_{z}^\alpha, Q_{x}^\beta, Q_{y}^\beta, Q_{z}^\beta \\ 
 &        & T_x, T_y, T_z & T_{u},T_{v},T_{yz},T_{zx},T_{xy} & T_{xyz}, T_{x}^\alpha, T_{y}^\alpha, T_{z}^\alpha, T_{x}^\beta, T_{y}^\beta, T_{z}^\beta \\ \hline
p&  & Q_0 & Q_x, Q_y, Q_z & Q_{u},Q_{v},Q_{yz},Q_{zx},Q_{xy}\\ 
  & & M_x, M_y, M_z & T_x, T_y, T_z &T_{u},T_{v},T_{yz},T_{zx},T_{xy} \\ 
  & & Q_{u}, Q_{v}, Q_{yz}, Q_{zx}, Q_{xy} & M_{u}, M_{v}, M_{yz}, M_{zx}, M_{xy} &M_{xyz},M_{x}^{\alpha},M_{y}^{\alpha},M_{z}^{\alpha},M_{x}^{\beta},M_{y}^{\beta},M_{z}^{\beta} \\ 
  & &  & G_{u}, G_{v}, G_{yz}, G_{zx}, G_{xy} &G_{xyz},G_{x}^{\alpha},G_{y}^{\alpha},G_{z}^{\alpha},G_{x}^{\beta},G_{y}^{\beta},G_{z}^{\beta} \\ 
  & &  & Q_{xyz}, Q_{x}^\alpha, Q_{y}^\alpha, Q_{z}^\alpha, Q_{x}^\beta, Q_{y}^\beta, Q_{z}^\beta &\text{E-hexadecapole (9)} \\ 
  & &  & T_{xyz}, T_{x}^\alpha, T_{y}^\alpha, T_{z}^\alpha, T_{x}^\beta, T_{y}^\beta, T_{z}^\beta &\text{MT-hexadecapole (9)} \\ \hline
d&  & & Q_0 & Q_x, Q_y, Q_z \\ 
  & & & M_x, M_y, M_z &  T_x, T_y, T_z\\ 
  & & & Q_{u}, Q_{v}, Q_{yz}, Q_{zx}, Q_{xy} & M_{u}, M_{v}, M_{yz}, M_{zx}, M_{xy} \\ 
  & & & M_{xyz}, M_x^\alpha, M_y^\alpha, M_z^\alpha, M_x^\beta, M_y^\beta, M_z^\beta  &G_{u}, G_{v}, G_{yz}, G_{zx}, G_{xy} \\ 
  & & & \text{E-hexadecapole (9)} &Q_{xyz}, Q_{x}^\alpha, Q_{y}^\alpha, Q_{z}^\alpha, Q_{x}^\beta, Q_{y}^\beta, Q_{z}^\beta  \\
   & & & & T_{xyz}, T_{x}^\alpha, T_{y}^\alpha, T_{z}^\alpha, T_{x}^\beta, T_{y}^\beta, T_{z}^\beta \\
    & & & & \text{M-hexadecapole (9)} \\
    & & & & \text{ET-hexadecapole (9)} \\
    & & & & \text{E-dotriacontapole (11)} \\
    & & & & \text{MT-dotriacontapole (11)} \\
    \hline
f& & & & Q_0 \\ 
  & & & & M_x, M_y, M_z \\ 
  & & & & Q_{u}, Q_{v}, Q_{yz}, Q_{zx}, Q_{xy} \\ 
  & & & & M_{xyz}, M_x^\alpha, M_y^\alpha, M_z^\alpha, M_x^\beta, M_y^\beta, M_z^\beta  \\ 
  & & & & \text{E-hexadecapole (9)} \\ 
  & & & & \text{M-dotriacontapole (11)} \\ 
  & & & & \text{E-tetrahexacontapole (13)} \\ \hline
\end{array}
\end{align}

\newpage
\section{Wave functions under toroidal multipole orderings}

In order to visualize wave functions under toroidal multipole orderings, we define the thermal average of an arbitrary operator $\hat{A}$,
\begin{align}
&
\braket{A(\hat{\bm{r}})}=\frac{1}{Z}\sum_{\alpha}e^{-\beta E_{\alpha}}{\rm Re}\left(\psi_{\alpha}^{*}(\hat{\bm{r}})\hat{A}\psi_{\alpha}^{}(\hat{\bm{r}})\right),
\quad
Z=\sum_{\alpha}e^{-\beta E_{\alpha}},
\\&
\braket{A}=\int d\hat{\bm{r}}\,\braket{A(\hat{\bm{r}})},
\end{align}
where $E_{\alpha}$ and $\psi_{\alpha}(\hat{\bm{r}})$ are the eigenenergies and corresponding wave functions, respectively.
In what follows, the shape and colormap of the wave function represent the electric charge density $\braket{A(\hat{\bm{r}})}$ with $\hat{A}=\hat{1}$ and the $z$-component of the orbital angular-momentum density $\braket{l^{z}(\hat{\bm{r}})}$, respectively.
We show specific three examples below at $1/\beta=0.1$.

\vspace*{2cm}
\subsection{Magnetic toroidal dipole in an $s$-$p$ hybridized-orbital system}
\begin{figure}[hbt!]
\begin{center}
\includegraphics[width=0.6 \hsize]{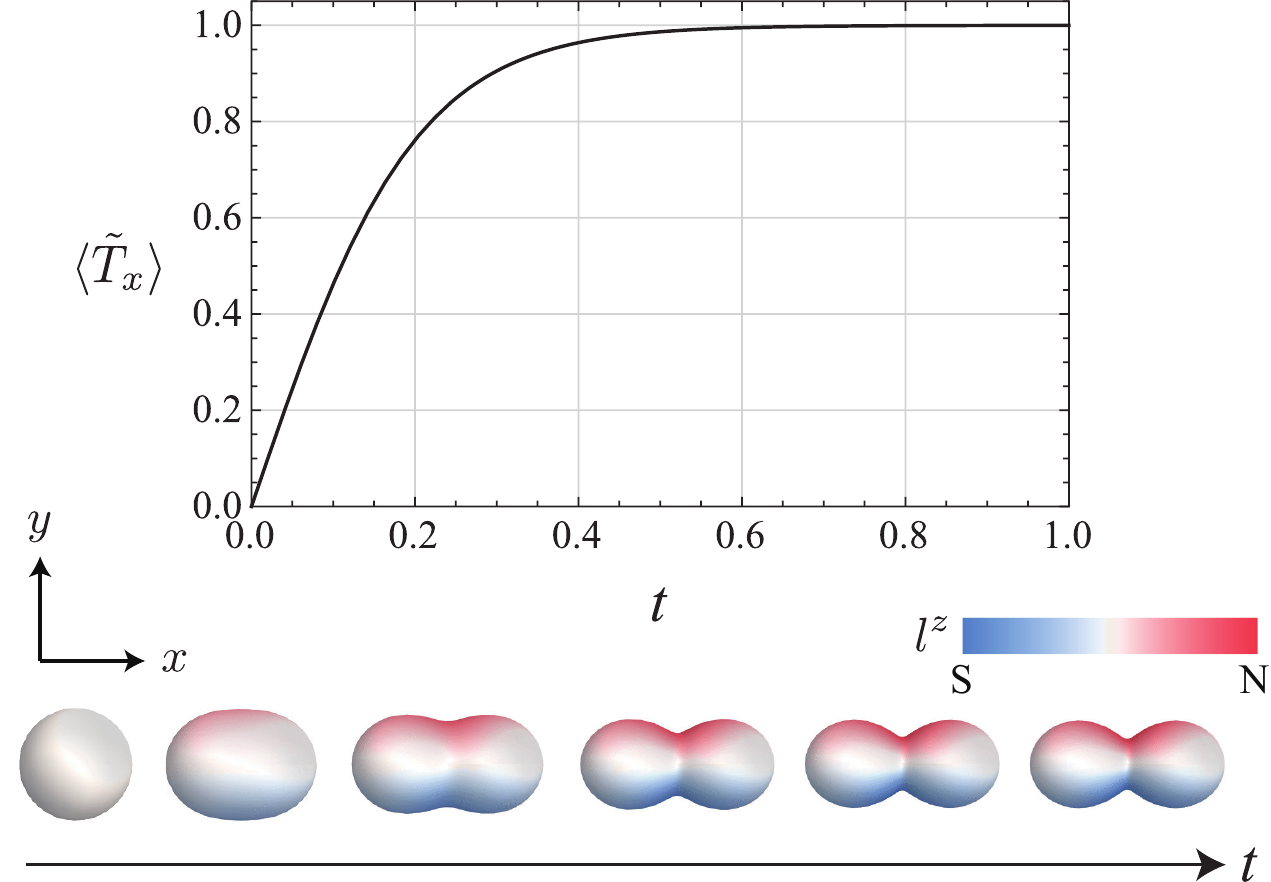} 
\caption{
\label{Fig:ME_sp_Temp}
(Upper panel) $t$ dependence of $\langle \tilde{T}_x \rangle$. 
(Lower panel) wave functions viewed from [001] for the $T_x$ ordering.  
}
\end{center}
\end{figure}

The Hamiltonian inducing the $T_x$ order in an $s$-$p$ hybridized-orbital system is given by 
\begin{align}
\mathcal{H}=-t \tilde{T}_x,  
\end{align}
where $t$ is the constant coefficient and $\tilde{T}_x=3\sqrt{3}T_x$. 
Under the basis wave function $(\phi_0,\phi_x,\phi_y,\phi_z)$, a thermal average of $\tilde{T}_x$ is shown in Fig.~\ref{Fig:ME_sp_Temp}. 
The wave functions are shown in the lower panel of Fig.~\ref{Fig:ME_sp_Temp}.

\vspace*{2cm}
\subsection{Magnetic toroidal quadrupole in an $s$-$d$ hybridized-orbital system}
\begin{figure}[hbt!]
\begin{center}
\includegraphics[width=0.6 \hsize]{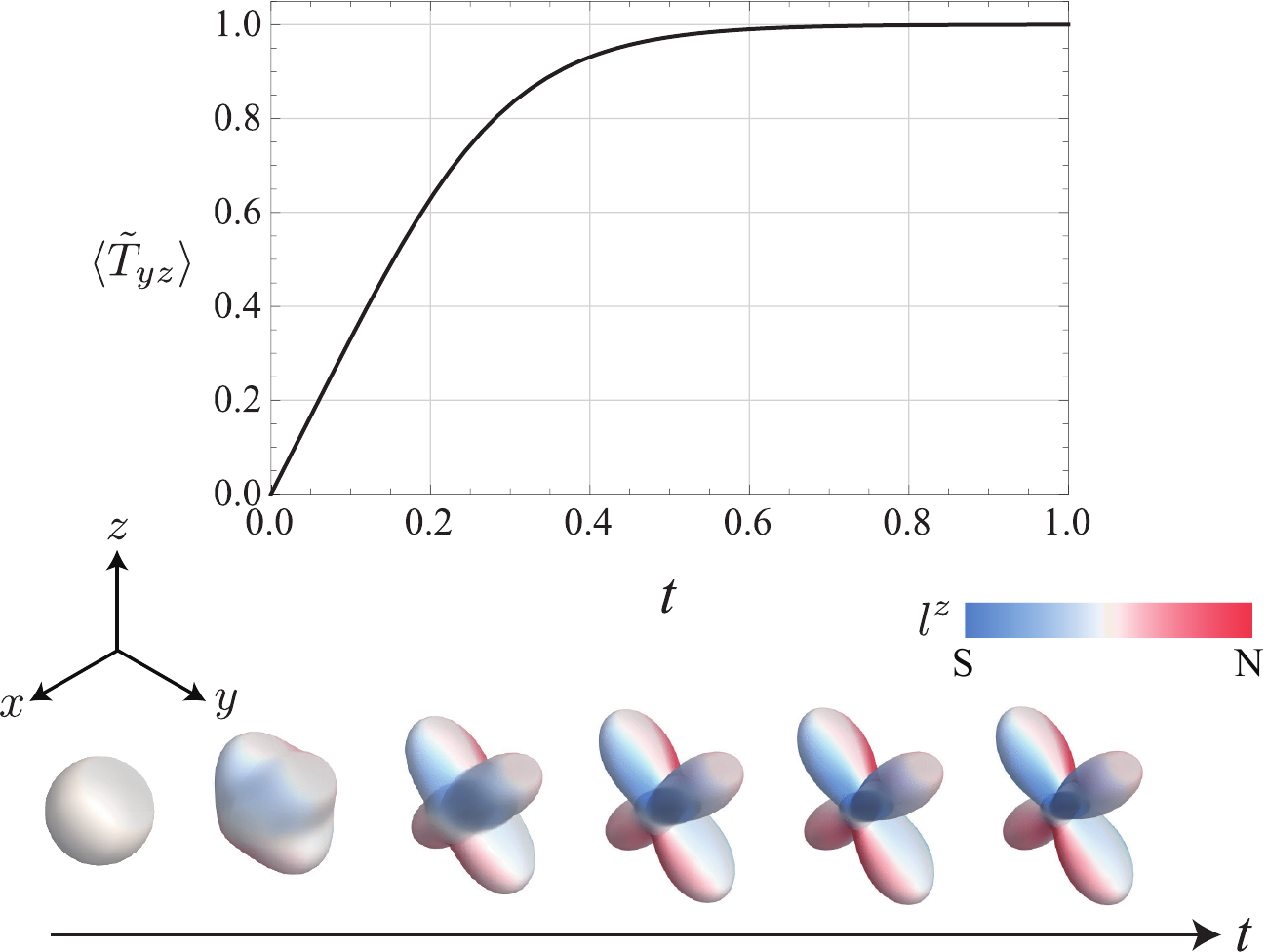} 
\caption{
\label{Fig:ME_sd_Temp}
(Upper panel) $t$ dependence of $\langle \tilde{T}_{yz} \rangle$. 
(Lower panel) wave functions viewed from [111] for the $T_{yz}$  ordering.  
}
\end{center}
\end{figure}
The Hamiltonian inducing the $T_{yz}$ order in an $s$-$d$ hybridized-orbital system is given by 
\begin{align}
\mathcal{H}=-t \tilde{T}_{yz}, 
\end{align}
where $t$ is the constant coefficient and $\tilde{T}_{yz}=2\sqrt{5}T_{yz}$. 
Under the basis wave function $(\phi_0,\phi_{u},\phi_{v},\phi_{yz},\phi_{zx},\phi_{xy})$, a thermal average of $\tilde{T}_{yz}$ is shown in Fig.~\ref{Fig:ME_sd_Temp}. 
The wave functions are shown in the lower panel of Fig.~\ref{Fig:ME_sd_Temp}.

\vspace*{2cm}
\subsection{Electric toroidal quadrupole in a $p$-$d$ hybridized-orbital system}
\begin{figure}[hbt!]
\begin{center}
\includegraphics[width=0.6 \hsize]{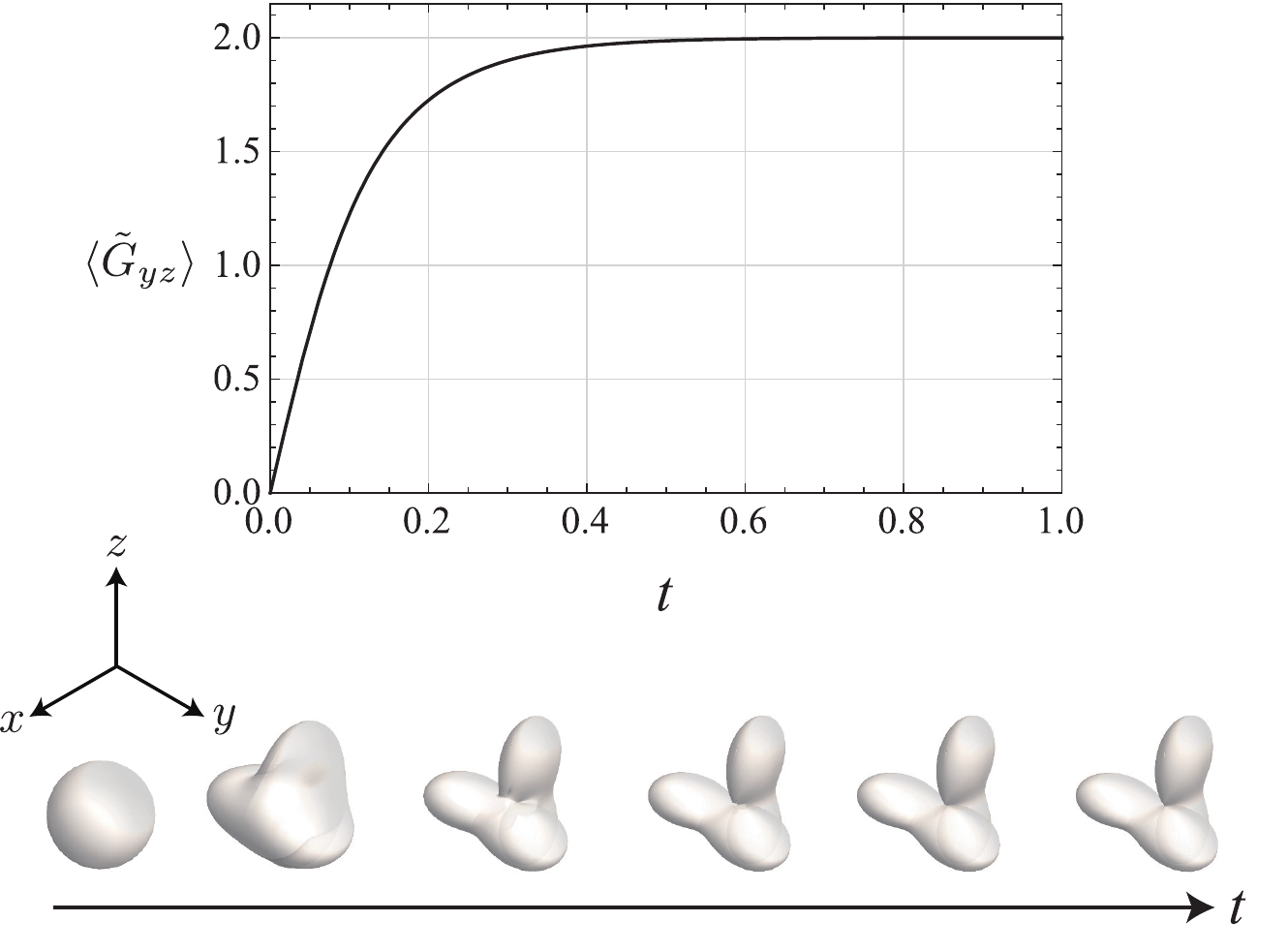} 
\caption{
\label{Fig:ME_pd1_Temp}
(Upper panel) $t$ dependence of $\langle \tilde{G}_{yz} \rangle$. 
(Lower panel) wave functions viewed from [111] for the $G_{yz}$  ordering.  
}
\end{center}
\end{figure}

The Hamiltonian inducing the $G_{yz}$ order in a $p$-$d$ hybridized-orbital system is given by 
\begin{align}
\mathcal{H}=-t \tilde{G}_{yz}, 
\end{align}
where $t$ is the constant coefficient and $\tilde{G}_{yz}=(3\sqrt{15}/2)G_{yz}$. 
Under the basis wave function $(\phi_x, \phi_y,\phi_z,\phi_{u},\phi_{v},\phi_{yz},\phi_{zx},\phi_{xy})$, a thermal average of $\tilde{G}_{yz}$ is shown in Fig.~\ref{Fig:ME_pd1_Temp}. 
The wave functions are shown in the lower panel of Fig.~\ref{Fig:ME_pd1_Temp}.

\newpage
\quad
\newpage
\section{Cross-correlated couplings under toroidal multipole ordering}
In this section, we show typical four examples of cross-correlated couplings in applied external fields under certain toroidal multipole orders. 

\vspace*{2cm}
\subsection{Magneto-electric coupling under a magnetic toroidal dipole ordering in an $s$-$p$ hybridized-orbital system}
\begin{figure}[hbt!]
\begin{center}
\includegraphics[width=0.6 \hsize]{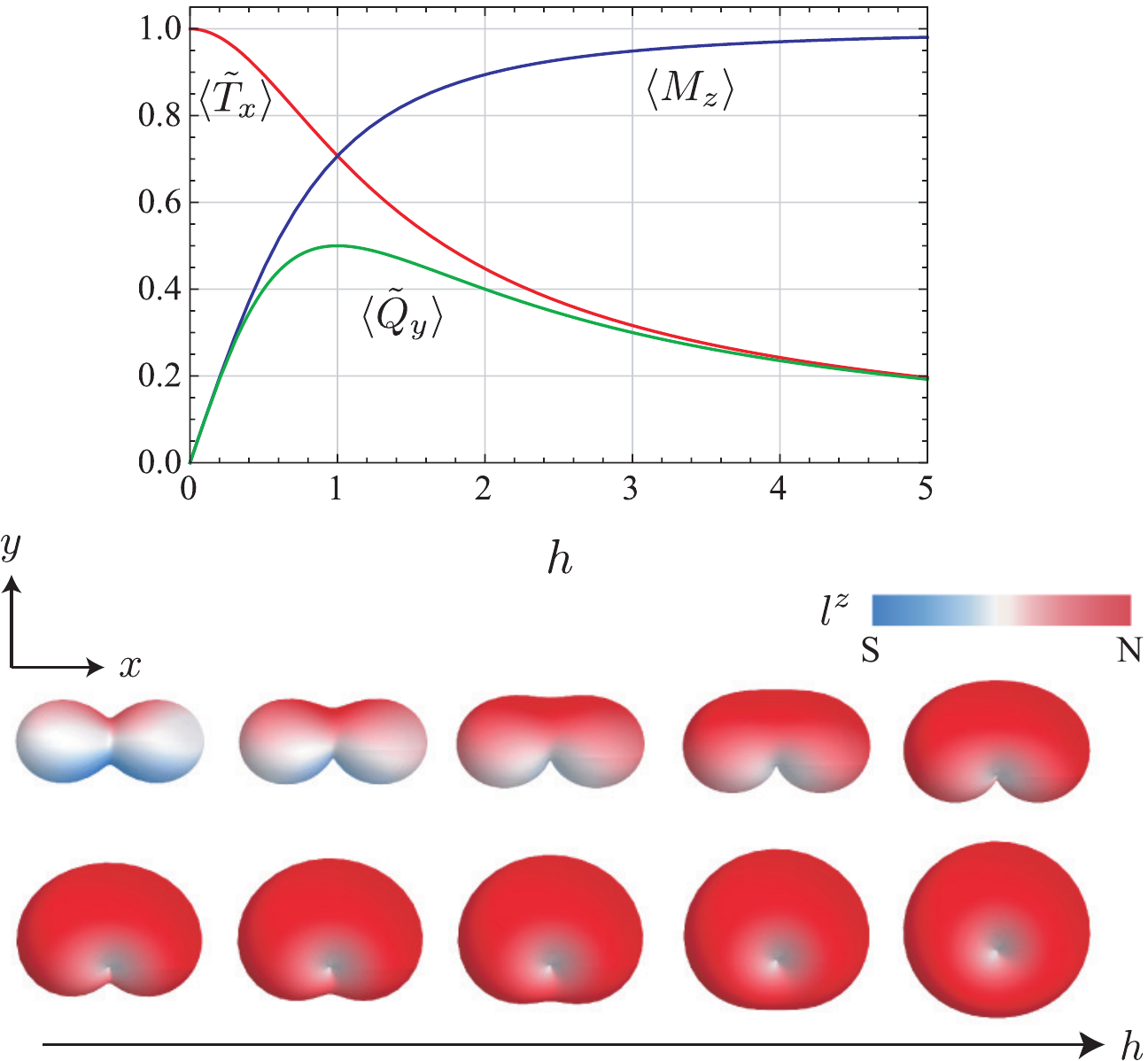} 
\caption{
\label{Fig:ME_sp_hdep}
(Upper panel) $h$ dependences of $\langle \tilde{T}_{x} \rangle$, $\langle M_{z} \rangle$, and $\langle \tilde{Q}_{y} \rangle$. 
(Lower panel) wave functions viewed from [001] modulated by external field.  
}
\end{center}
\end{figure}

We show magneto-electric coupling under a magnetic toroidal dipole ordering $T_x$ in an $s$-$p$ hybridized-orbital system. 
The Hamiltonian under external magnetic field is given by 
\begin{align}
\mathcal{H}=-\tilde{T}_x -h M_z, 
\end{align}
where $h$ is the constant coefficient. 
The second term is the coupling with external magnetic field along the $z$ direction. 
By taking a thermal average of $\tilde{Q}_{y}=\sqrt{3}Q_{y}$, we find that the electric polarization $Q_y$ is induced by external magnetic field $M_z$ under the magnetic toroidal dipole ordering $T_x$, which is so-called the magneto-electric effect.  
Figure~\ref{Fig:ME_sp_hdep} shows $h$ dependences of $\langle \tilde{T}_{x} \rangle$, $\langle M_{z} \rangle$, and $\langle \tilde{Q}_{y} \rangle$.
The wave functions are shown in the lower panel of Fig.~\ref{Fig:ME_sp_hdep}. 

\vspace*{2cm}
\subsection{Magneto-elastic coupling under a magnetic toroidal quadrupole ordering in an $s$-$d$ hybridized-orbital system}

\begin{figure}[hbt!]
\begin{center}
\includegraphics[width=0.6 \hsize]{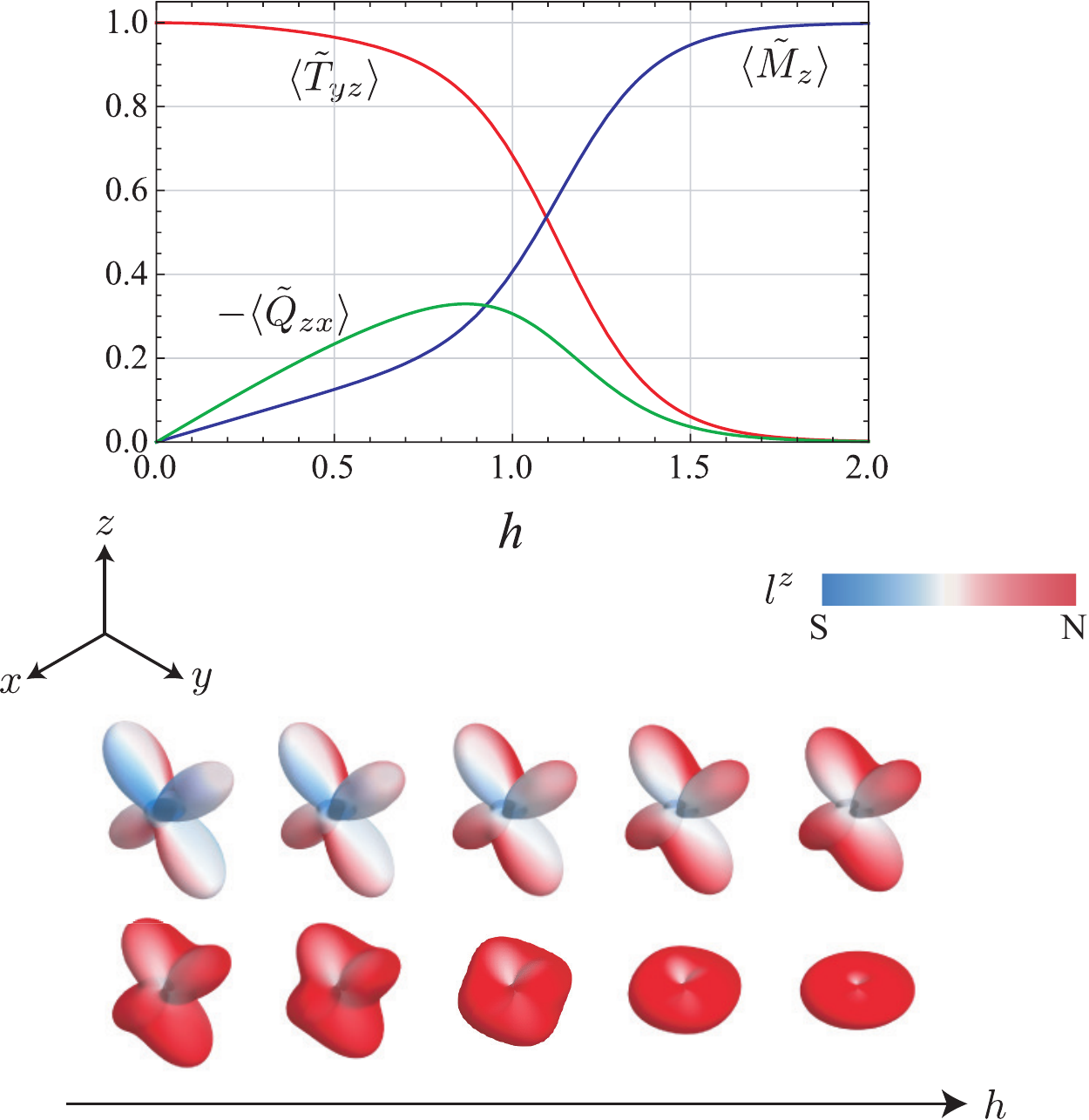} 
\caption{
\label{Fig:ME_sd_hdep}
(Upper panel) $h$ dependences of $\langle \tilde{T}_{yz} \rangle$, $\langle \tilde{M}_{z} \rangle$, and $\langle \tilde{Q}_{zx} \rangle$. 
(Lower panel) wave functions viewed from [111] modulated by external field.  
}
\end{center}
\end{figure}

We show magneto-elastic coupling under a magnetic toroidal quadrupole ordering $T_{yz}$ in an $s$-$d$ hybridized-orbital system. 
The Hamiltonian under external magnetic field is given by 
\begin{align}
\mathcal{H}=-\tilde{T}_{yz} -h \tilde{M}_z, 
\end{align}
where $h$ is the constant coefficient and $\tilde{M}_z=M_z/2$. 
The second term is the coupling with external magnetic field along the $z$ direction. 
By taking a thermal average of $\tilde{Q}_{zx}=\sqrt{5}Q_{zx}$, we find that the quadrupole-type distortion $Q_{zx}$ is induced by external magnetic field $M_z$ under the magnetic toroidal quadrupole ordering $T_{yz}$, which is so-called the magneto-elastic effect.  
Figure~\ref{Fig:ME_sd_hdep} shows $h$ dependences of $\langle \tilde{T}_{yz} \rangle$, $\langle \tilde{M}_{z} \rangle$, and $\langle \tilde{Q}_{zx} \rangle$.
The wave functions are shown in the lower panel of Fig.~\ref{Fig:ME_sd_hdep}. 

\vspace*{2cm}
\subsection{Magneto-toroidal coupling under an electric toroidal quadrupole ordering in a $p$-$d$ hybridized-orbital system}

\begin{figure}[hbt!]
\begin{center}
\includegraphics[width=0.6 \hsize]{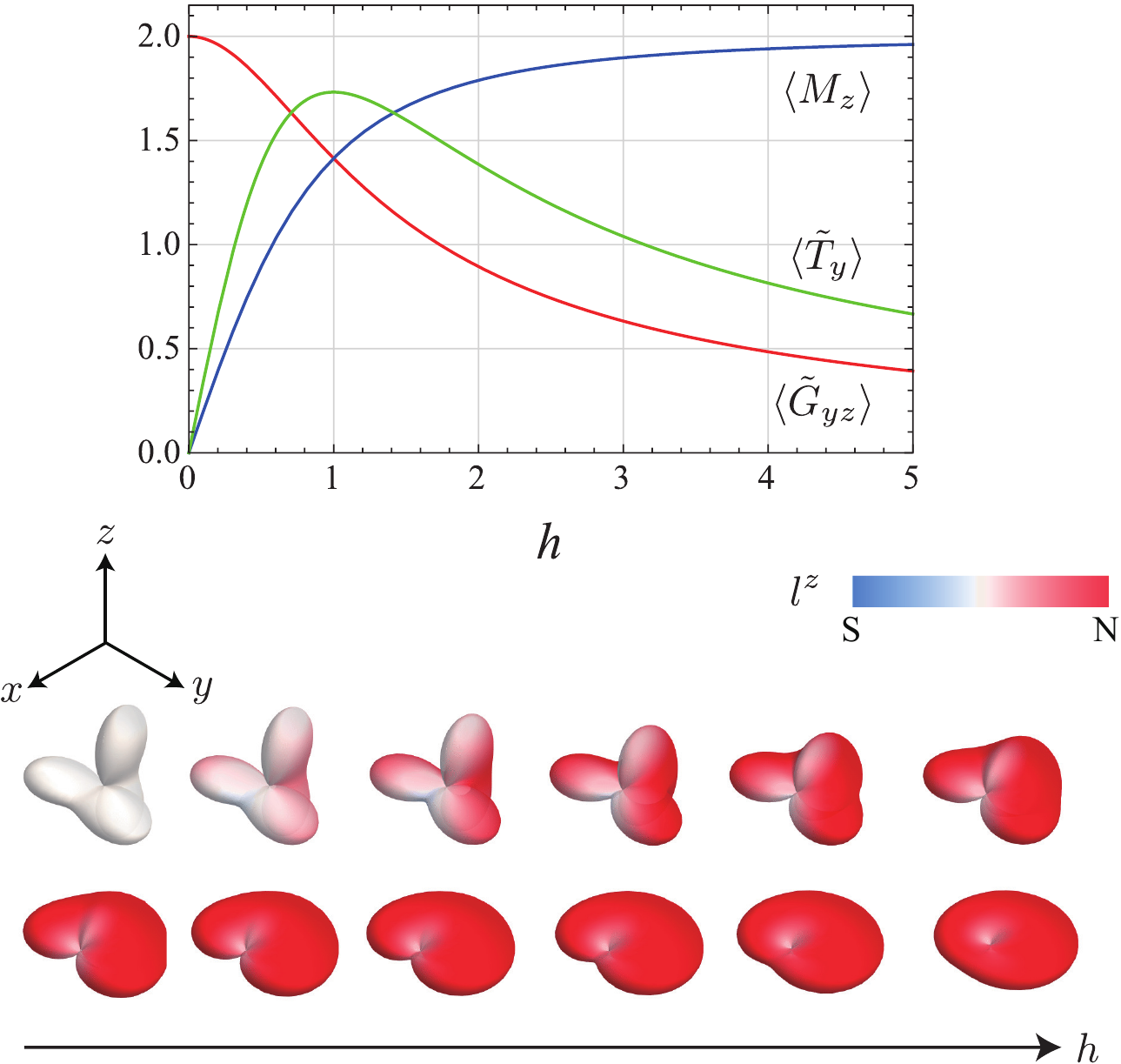} 
\caption{
\label{Fig:ME_pd1_hdep}]
(Upper panel) $h$ dependences of $\langle \tilde{G}_{yz} \rangle$, $\langle M_{z} \rangle$, and $\langle \tilde{T}_{y} \rangle$. 
(Lower panel) wave functions viewed from [111] modulated by external field.  
}
\end{center}
\end{figure}

We show an unconventional coupling between magnetic and magnetic toroidal moments under an electric toroidal quadrupole ordering $G_{yz}$ in a $p$-$d$ hybridized-orbital system. 
The Hamiltonian under external magnetic field is given by 
\begin{align}
\mathcal{H}=-\tilde{G}_{yz} -h M_z, 
\end{align}
where $h$ is the constant coefficient. 
The second term is the coupling with external magnetic field along the $z$ direction. 
By taking a thermal average of $\tilde{T}_{y}=(3\sqrt{15}/2)T_{y}$, we find that the magnetic toroidal dipole $T_{y}$ is induced by external magnetic field $M_z$ under the electric toroidal quadrupole ordering $G_{yz}$, indicating the coupling between the magnetic dipole and magnetic toroidal moments.  
Figure~\ref{Fig:ME_pd1_hdep} shows $h$ dependences of $\langle \tilde{G}_{yz} \rangle$, $\langle M_{z} \rangle$, and $\langle \tilde{T}_{y} \rangle$.
The wave functions are shown in the lower panel of Fig.~\ref{Fig:ME_pd1_hdep}.

\vspace*{2cm}
\subsection{Electro-elastic coupling under an electric toroidal quadrupole ordering in a $p$-$d$ hybridized-orbital system}

\begin{figure}[hbt!]
\begin{center}
\includegraphics[width=0.6 \hsize]{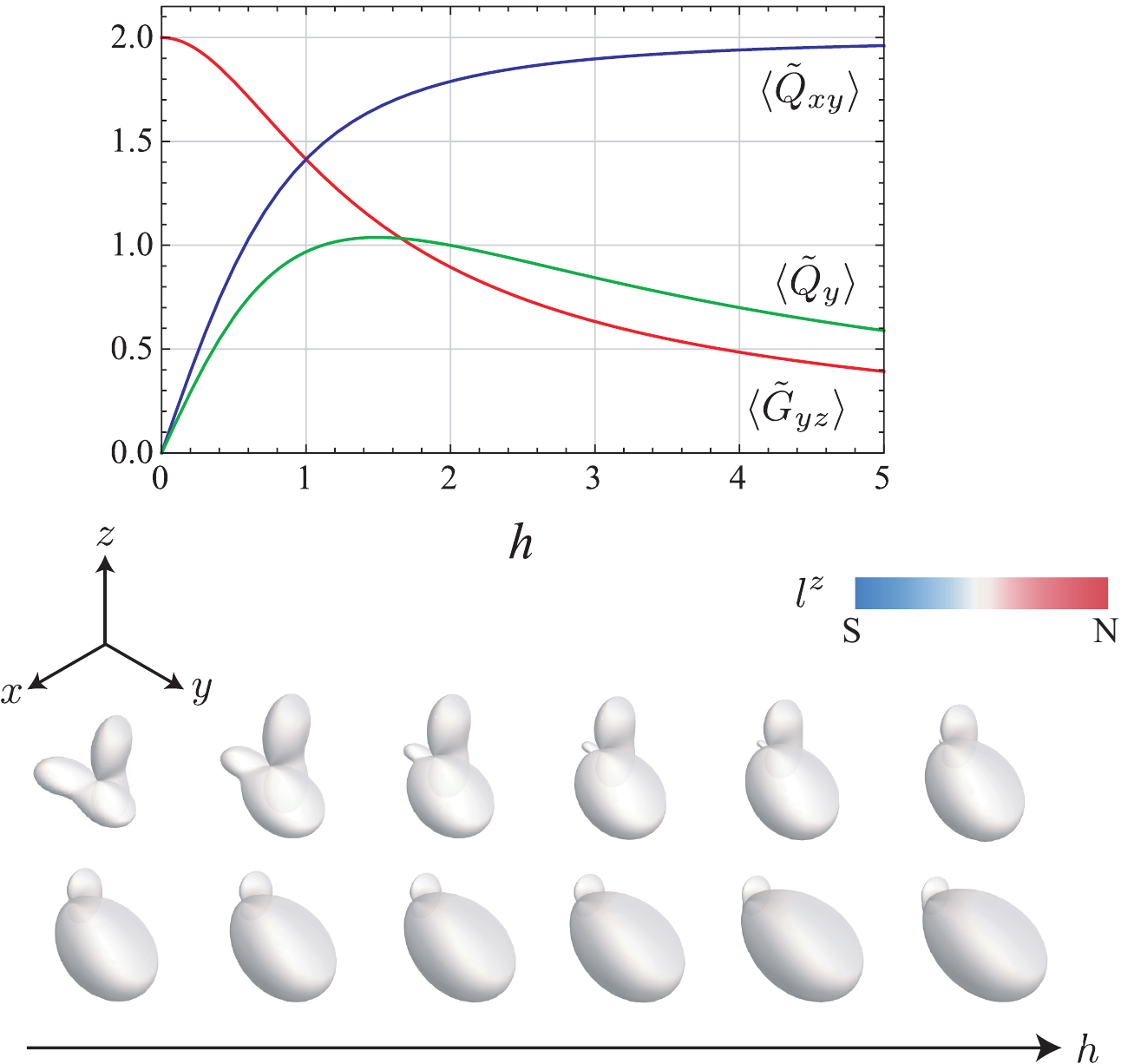} 
\caption{
\label{Fig:ME_pd2_hdep}
(Upper panel) $h$ dependences of $\langle \tilde{G}_{yz} \rangle$, $\langle \tilde{Q}_{y} \rangle$, and $\langle \tilde{Q}_{xy} \rangle$. 
(Lower panel) wave functions viewed from [111] modulated by external field.  
}
\end{center}
\end{figure}

We show electro-elastic coupling under an electric toroidal quadrupole ordering $G_{yz}$ in a $d$-$p$ hybridized-orbital system. 
The Hamiltonian under external electric field is given by 
\begin{align}
\mathcal{H}=-\tilde{G}_{yz} -h \tilde{Q}_y, 
\end{align}
where $h$ is the constant coefficient and $\tilde{Q}_y=\sqrt{15}Q_y$. 
The second term is the coupling with external electric field along the $y$ direction. 
By taking a thermal average of $\tilde{Q}_{xy}=(35/8)Q_{xy}$, we find that the quadrupole-type distortion $Q_{xy}$ is induced by external electric field $Q_y$ under the electric toroidal quadrupole ordering $G_{yz}$, which is so-called the electro-elastic effect.  
Figure~\ref{Fig:ME_pd2_hdep} shows $h$ dependences of $\langle \tilde{G}_{yz} \rangle$, $\langle Q_{y} \rangle$, and $\langle \tilde{Q}_{xy} \rangle$.
The wave functions are shown in the lower panel of Fig.~\ref{Fig:ME_pd2_hdep}.

\newpage
\section{Matrix elements of hybrid mutlipoles}
In this section, we explicitly show matrix elements of multipoles in each hybridized-orbital system. 
The basis wave functions for $s$, $p$, $d$, and $f$ orbitals as a function of angles $\hat{\bm{r}}/r$ are given by 
\begin{align}
\underline{s\,(L=0)}&
\cr&
\phi_{0}=\frac{1}{\sqrt{4\pi}},
\\
\underline{p\,(L=1)}&
\cr&
\phi_{x}=\sqrt{\frac{3}{4\pi}}\frac{x}{r},
\quad
\phi_{y}=\sqrt{\frac{3}{4\pi}}\frac{y}{r},
\quad
\phi_{z}=\sqrt{\frac{3}{4\pi}}\frac{z}{r},
\\
\underline{d\,(L=2)}&
\cr&
\phi_{u}=\sqrt{\frac{5}{4\pi}}\frac{1}{2}\frac{3z^{2}-r^{2}}{r^2} ,
\quad
\phi_{v}=\sqrt{\frac{5}{4\pi}}\frac{\sqrt{3}}{2}\frac{x^{2}-y^{2}}{r^2},
\quad
\phi_{yz}=\sqrt{\frac{5}{4\pi}}\sqrt{3}\frac{yz}{r^2},
\cr&
\phi_{zx}=\sqrt{\frac{5}{4\pi}}\sqrt{3}\frac{zx}{r^2},
\quad
\phi_{xy}=\sqrt{\frac{5}{4\pi}}\sqrt{3}\frac{xy}{r^2},
\\
\underline{f\,(L=3)}&
\cr&
\phi_{xyz}=\sqrt{\frac{7}{4\pi}}\sqrt{15}\frac{xyz}{r^3},
\quad
\phi_{x}^{\alpha}=\sqrt{\frac{7}{4\pi}}\frac{1}{2}\frac{x(5x^{2}-3r^{2})}{r^3},
\cr&
\phi_{y}^{\alpha}=\sqrt{\frac{7}{4\pi}}\frac{1}{2}\frac{y(5y^{2}-3r^{2})}{r^3},
\quad
\phi_{z}^{\alpha}=\sqrt{\frac{7}{4\pi}}\frac{1}{2}\frac{z(5z^{2}-3r^{2})}{r^3},
\cr&
\phi_{x}^{\beta}=\sqrt{\frac{7}{4\pi}}\frac{\sqrt{15}}{2}\frac{x(y^{2}-z^{2})}{r^3},
\quad
\phi_{y}^{\beta}=\sqrt{\frac{7}{4\pi}}\frac{\sqrt{15}}{2}\frac{y(z^{2}-x^{2})}{r^3},
\cr&
\phi_{z}^{\beta}=\sqrt{\frac{7}{4\pi}}\frac{\sqrt{15}}{2}\frac{z(x^{2}-y^{2})}{r^3}.
\end{align}
We show each situation one by one below. 

\subsection{$s$-$p$ hybridized-orbital system}

Matrix elements of each multipole activated in the $s$-$p$ hybridized-orbital system are given below. 
The basis function is taken as $(\phi_0,\phi_x,\phi_y,\phi_z)$. 

\begin{align}
&
Q_{x}=\frac{1}{\sqrt{3}}
\left(
\begin{array}{c|ccc}
 0 & 1 & 0 & 0 \\ \hline
 1 & 0 & 0 & 0 \\
 0 & 0 & 0 & 0 \\
 0 & 0 & 0 & 0 \\
\end{array}
\right),
\ \
Q_{y}=\frac{1}{\sqrt{3}}
\left(
\begin{array}{c|ccc}
 0 & 0 & 1 & 0 \\ \hline
 0 & 0 & 0 & 0 \\
 1 & 0 & 0 & 0 \\
 0 & 0 & 0 & 0 \\
\end{array}
\right),
\ \
Q_{z}=\frac{1}{\sqrt{3}}
\left(
\begin{array}{c|ccc}
 0 & 0 & 0 & 1 \\ \hline
 0 & 0 & 0 & 0 \\
 0 & 0 & 0 & 0 \\
 1 & 0 & 0 & 0 \\
\end{array}
\right),
\\&
T_{x}=\frac{1}{3\sqrt{3}}
\left(
\begin{array}{c|ccc}
 0 & i & 0 & 0 \\ \hline
 -i & 0 & 0 & 0 \\
 0 & 0 & 0 & 0 \\
 0 & 0 & 0 & 0 \\
\end{array}
\right),
\ \ 
T_{y}=\frac{1}{3\sqrt{3}}
\left(
\begin{array}{c|ccc}
 0 & 0 & i & 0 \\ \hline
 0 & 0 & 0 & 0 \\
 -i & 0 & 0 & 0 \\
 0 & 0 & 0 & 0 \\
\end{array}
\right),
\ \ 
T_{z}=\frac{1}{3\sqrt{3}}
\left(
\begin{array}{c|ccc}
 0 & 0 & 0 & i \\ \hline
 0 & 0 & 0 & 0 \\
 0 & 0 & 0 & 0 \\
 -i & 0 & 0 & 0 \\
\end{array}
\right),
\\&
M_{x}=
\left(
\begin{array}{c|ccc}
 0 & 0 & 0 & 0 \\ \hline
 0 & 0 & 0 & 0 \\
 0 & 0 & 0 & -i \\
 0 & 0 & i & 0 \\
\end{array}
\right),
\ \
M_{y}=
\left(
\begin{array}{c|ccc}
 0 & 0 & 0 & 0 \\ \hline
 0 & 0 & 0 & i \\
 0 & 0 & 0 & 0 \\
 0 & -i & 0 & 0 \\
\end{array}
\right),
\ \ 
M_{z}=
\left(
\begin{array}{c|ccc}
 0 & 0 & 0 & 0 \\ \hline
 0 & 0 & -i & 0 \\
 0 & i & 0 & 0 \\
 0 & 0 & 0 & 0 \\
\end{array}
\right),
\\ &
Q_{u}=\frac{1}{5}
\left(
\begin{array}{c|ccc}
 0 & 0 & 0 & 0 \\ \hline
 0 & -1 & 0 & 0 \\
 0 & 0 & -1 & 0 \\
 0 & 0 & 0 & 2 \\
\end{array}
\right),
\ \
Q_{v}=\frac{\sqrt{3}}{5}
\left(
\begin{array}{c|ccc}
 0 & 0 & 0 & 0 \\ \hline
 0 & 1 & 0 & 0 \\
 0 & 0 & -1 & 0 \\
 0 & 0 & 0 & 0 \\
\end{array}
\right),
\nonumber \\&
Q_{yz}=\frac{\sqrt{3}}{5}
\left(
\begin{array}{c|ccc}
 0 & 0 & 0 & 0 \\ \hline
 0 & 0 & 0 & 0 \\
 0 & 0 & 0 & 1 \\
 0 & 0 & 1 & 0 \\
\end{array}
\right),
\ \ 
Q_{zx}=\frac{\sqrt{3}}{5}
\left(
\begin{array}{c|ccc}
 0 & 0 & 0 & 0 \\ \hline
 0 & 0 & 0 & 1 \\
 0 & 0 & 0 & 0 \\
 0 & 1 & 0 & 0 \\
\end{array}
\right),
\ \ 
Q_{xy}=\frac{\sqrt{3}}{5}
\left(
\begin{array}{c|ccc}
 0 & 0 & 0 & 0 \\ \hline
 0 & 0 & 1 & 0 \\
 0 & 1 & 0 & 0 \\
 0 & 0 & 0 & 0 \\
\end{array}
\right).
\end{align}

\subsection{$s$-$f$ hybridized-orbital system}
Matrix elements of each multipole activated in the $s$-$f$ hybridized-orbital system are given below. 
The basis function is taken as $(\phi_0,\phi_{xyz},\phi_{x}^{\alpha},\phi_{y}^{\alpha},\phi_{z}^{\alpha},\phi_{x}^{\beta},\phi_{y}^{\beta},\phi_{z}^{\beta})$. 

\begin{align}
&
Q_{xyz}=\frac{1}{\sqrt{7}}
\left(
\begin{array}{c|ccccccc}
 0 & 1 & 0 & 0 & 0 & 0 & 0 & 0 \\ \hline
 1 & 0 & 0 & 0 & 0 & 0 & 0 & 0 \\
 0 & 0 & 0 & 0 & 0 & 0 & 0 & 0 \\
 0 & 0 & 0 & 0 & 0 & 0 & 0 & 0 \\
 0 & 0 & 0 & 0 & 0 & 0 & 0 & 0 \\
 0 & 0 & 0 & 0 & 0 & 0 & 0 & 0 \\
 0 & 0 & 0 & 0 & 0 & 0 & 0 & 0 \\
 0 & 0 & 0 & 0 & 0 & 0 & 0 & 0 \\
\end{array}
\right),
\cr&
Q^{\alpha}_{x}=\frac{1}{\sqrt{7}}
\left(
\begin{array}{c|ccccccc}
 0 & 0 & 1 & 0 & 0 & 0 & 0 & 0 \\ \hline
 0 & 0 & 0 & 0 & 0 & 0 & 0 & 0 \\
 1 & 0 & 0 & 0 & 0 & 0 & 0 & 0 \\
 0 & 0 & 0 & 0 & 0 & 0 & 0 & 0 \\
 0 & 0 & 0 & 0 & 0 & 0 & 0 & 0 \\
 0 & 0 & 0 & 0 & 0 & 0 & 0 & 0 \\
 0 & 0 & 0 & 0 & 0 & 0 & 0 & 0 \\
 0 & 0 & 0 & 0 & 0 & 0 & 0 & 0 \\
\end{array}
\right),
Q^{\alpha}_{y}=\frac{1}{\sqrt{7}}
\left(
\begin{array}{c|ccccccc}
 0 & 0 & 0 & 1 & 0 & 0 & 0 & 0 \\ \hline
 0 & 0 & 0 & 0 & 0 & 0 & 0 & 0 \\
 0 & 0 & 0 & 0 & 0 & 0 & 0 & 0 \\
 1 & 0 & 0 & 0 & 0 & 0 & 0 & 0 \\
 0 & 0 & 0 & 0 & 0 & 0 & 0 & 0 \\
 0 & 0 & 0 & 0 & 0 & 0 & 0 & 0 \\
 0 & 0 & 0 & 0 & 0 & 0 & 0 & 0 \\
 0 & 0 & 0 & 0 & 0 & 0 & 0 & 0 \\
\end{array}
\right),
Q^{\alpha}_{z}=\frac{1}{\sqrt{7}}
\left(
\begin{array}{c|ccccccc}
 0 & 0 & 0 & 0 & 1 & 0 & 0 & 0 \\ \hline
 0 & 0 & 0 & 0 & 0 & 0 & 0 & 0 \\
 0 & 0 & 0 & 0 & 0 & 0 & 0 & 0 \\
 0 & 0 & 0 & 0 & 0 & 0 & 0 & 0 \\
 1 & 0 & 0 & 0 & 0 & 0 & 0 & 0 \\
 0 & 0 & 0 & 0 & 0 & 0 & 0 & 0 \\
 0 & 0 & 0 & 0 & 0 & 0 & 0 & 0 \\
 0 & 0 & 0 & 0 & 0 & 0 & 0 & 0 \\
\end{array}
\right),
\cr&
Q^{\beta}_{x}=\frac{1}{\sqrt{7}}
\left(
\begin{array}{c|ccccccc}
 0 & 0 & 0 & 0 & 0 & 1 & 0 & 0 \\ \hline
 0 & 0 & 0 & 0 & 0 & 0 & 0 & 0 \\
 0 & 0 & 0 & 0 & 0 & 0 & 0 & 0 \\
 0 & 0 & 0 & 0 & 0 & 0 & 0 & 0 \\
 0 & 0 & 0 & 0 & 0 & 0 & 0 & 0 \\
 1 & 0 & 0 & 0 & 0 & 0 & 0 & 0 \\
 0 & 0 & 0 & 0 & 0 & 0 & 0 & 0 \\
 0 & 0 & 0 & 0 & 0 & 0 & 0 & 0 \\
\end{array}
\right),
Q^{\beta}_{y}=\frac{1}{\sqrt{7}}
\left(
\begin{array}{c|ccccccc}
 0 & 0 & 0 & 0 & 0 & 0 & 1 & 0 \\ \hline
 0 & 0 & 0 & 0 & 0 & 0 & 0 & 0 \\
 0 & 0 & 0 & 0 & 0 & 0 & 0 & 0 \\
 0 & 0 & 0 & 0 & 0 & 0 & 0 & 0 \\
 0 & 0 & 0 & 0 & 0 & 0 & 0 & 0 \\
 0 & 0 & 0 & 0 & 0 & 0 & 0 & 0 \\
 1 & 0 & 0 & 0 & 0 & 0 & 0 & 0 \\
 0 & 0 & 0 & 0 & 0 & 0 & 0 & 0 \\
\end{array}
\right),
Q^{\beta}_{z}=\frac{1}{\sqrt{7}}
\left(
\begin{array}{c|ccccccc}
 0 & 0 & 0 & 0 & 0 & 0 & 0 & 1 \\ \hline
 0 & 0 & 0 & 0 & 0 & 0 & 0 & 0 \\
 0 & 0 & 0 & 0 & 0 & 0 & 0 & 0 \\
 0 & 0 & 0 & 0 & 0 & 0 & 0 & 0 \\
 0 & 0 & 0 & 0 & 0 & 0 & 0 & 0 \\
 0 & 0 & 0 & 0 & 0 & 0 & 0 & 0 \\
 0 & 0 & 0 & 0 & 0 & 0 & 0 & 0 \\
 1 & 0 & 0 & 0 & 0 & 0 & 0 & 0 \\
\end{array}
\right), 
\\ &
T_{xyz}=\frac{3}{5\sqrt{7}}
\left(
\begin{array}{c|ccccccc}
 0 & i & 0 & 0 & 0 & 0 & 0 & 0 \\ \hline
 -i & 0 & 0 & 0 & 0 & 0 & 0 & 0 \\
 0 & 0 & 0 & 0 & 0 & 0 & 0 & 0 \\
 0 & 0 & 0 & 0 & 0 & 0 & 0 & 0 \\
 0 & 0 & 0 & 0 & 0 & 0 & 0 & 0 \\
 0 & 0 & 0 & 0 & 0 & 0 & 0 & 0 \\
 0 & 0 & 0 & 0 & 0 & 0 & 0 & 0 \\
 0 & 0 & 0 & 0 & 0 & 0 & 0 & 0 \\
\end{array}
\right),
\cr&
T^{\alpha}_{x}=\frac{3}{5\sqrt{7}}
\left(
\begin{array}{c|ccccccc}
 0 & 0 & i & 0 & 0 & 0 & 0 & 0 \\ \hline
 0 & 0 & 0 & 0 & 0 & 0 & 0 & 0 \\
 -i & 0 & 0 & 0 & 0 & 0 & 0 & 0 \\
 0 & 0 & 0 & 0 & 0 & 0 & 0 & 0 \\
 0 & 0 & 0 & 0 & 0 & 0 & 0 & 0 \\
 0 & 0 & 0 & 0 & 0 & 0 & 0 & 0 \\
 0 & 0 & 0 & 0 & 0 & 0 & 0 & 0 \\
 0 & 0 & 0 & 0 & 0 & 0 & 0 & 0 \\
\end{array}
\right),
T^{\alpha}_{y}=\frac{3}{5\sqrt{7}}
\left(
\begin{array}{c|ccccccc}
 0 & 0 & 0 & i & 0 & 0 & 0 & 0 \\ \hline
 0 & 0 & 0 & 0 & 0 & 0 & 0 & 0 \\
 0 & 0 & 0 & 0 & 0 & 0 & 0 & 0 \\
 -i & 0 & 0 & 0 & 0 & 0 & 0 & 0 \\
 0 & 0 & 0 & 0 & 0 & 0 & 0 & 0 \\
 0 & 0 & 0 & 0 & 0 & 0 & 0 & 0 \\
 0 & 0 & 0 & 0 & 0 & 0 & 0 & 0 \\
 0 & 0 & 0 & 0 & 0 & 0 & 0 & 0 \\
\end{array}
\right),
T^{\alpha}_{z}=\frac{3}{5\sqrt{7}}
\left(
\begin{array}{c|ccccccc}
 0 & 0 & 0 & 0 & i & 0 & 0 & 0 \\ \hline
 0 & 0 & 0 & 0 & 0 & 0 & 0 & 0 \\
 0 & 0 & 0 & 0 & 0 & 0 & 0 & 0 \\
 0 & 0 & 0 & 0 & 0 & 0 & 0 & 0 \\
 -i & 0 & 0 & 0 & 0 & 0 & 0 & 0 \\
 0 & 0 & 0 & 0 & 0 & 0 & 0 & 0 \\
 0 & 0 & 0 & 0 & 0 & 0 & 0 & 0 \\
 0 & 0 & 0 & 0 & 0 & 0 & 0 & 0 \\
\end{array}
\right),
\cr&
T^{\beta}_{x}=\frac{3}{5\sqrt{7}}
\left(
\begin{array}{c|ccccccc}
 0 & 0 & 0 & 0 & 0 & i & 0 & 0 \\ \hline
 0 & 0 & 0 & 0 & 0 & 0 & 0 & 0 \\
 0 & 0 & 0 & 0 & 0 & 0 & 0 & 0 \\
 0 & 0 & 0 & 0 & 0 & 0 & 0 & 0 \\
 0 & 0 & 0 & 0 & 0 & 0 & 0 & 0 \\
 -i & 0 & 0 & 0 & 0 & 0 & 0 & 0 \\
 0 & 0 & 0 & 0 & 0 & 0 & 0 & 0 \\
 0 & 0 & 0 & 0 & 0 & 0 & 0 & 0 \\
\end{array}
\right),
T^{\beta}_{y}=\frac{3}{5\sqrt{7}}
\left(
\begin{array}{c|ccccccc}
 0 & 0 & 0 & 0 & 0 & 0 & i & 0 \\ \hline
 0 & 0 & 0 & 0 & 0 & 0 & 0 & 0 \\
 0 & 0 & 0 & 0 & 0 & 0 & 0 & 0 \\
 0 & 0 & 0 & 0 & 0 & 0 & 0 & 0 \\
 0 & 0 & 0 & 0 & 0 & 0 & 0 & 0 \\
 0 & 0 & 0 & 0 & 0 & 0 & 0 & 0 \\
 -i & 0 & 0 & 0 & 0 & 0 & 0 & 0 \\
 0 & 0 & 0 & 0 & 0 & 0 & 0 & 0 \\
\end{array}
\right),
T^{\beta}_{z}=\frac{3}{5\sqrt{7}}
\left(
\begin{array}{c|ccccccc}
 0 & 0 & 0 & 0 & 0 & 0 & 0 & i \\ \hline
 0 & 0 & 0 & 0 & 0 & 0 & 0 & 0 \\
 0 & 0 & 0 & 0 & 0 & 0 & 0 & 0 \\
 0 & 0 & 0 & 0 & 0 & 0 & 0 & 0 \\
 0 & 0 & 0 & 0 & 0 & 0 & 0 & 0 \\
 0 & 0 & 0 & 0 & 0 & 0 & 0 & 0 \\
 0 & 0 & 0 & 0 & 0 & 0 & 0 & 0 \\
 -i & 0 & 0 & 0 & 0 & 0 & 0 & 0 \\
\end{array}
\right).
\cr
\end{align}

\subsection{$p$-$d$ hybridized-orbital system}
Matrix elements of each multipole activated in the $p$-$d$ hybridized-orbital system are given below. 
The basis function is taken as $(\phi_x, \phi_y,\phi_z,\phi_{u},\phi_{v},\phi_{yz},\phi_{zx},\phi_{xy})$. 

\begin{align}
&
Q_{x}=\frac{1}{\sqrt{15}}
\left(
\begin{array}{ccc|ccccc}
 0 & 0 & 0 & -1 & \sqrt{3} & 0 & 0 & 0 \\
 0 & 0 & 0 & 0 & 0 & 0 & 0 & \sqrt{3} \\
 0 & 0 & 0 & 0 & 0 & 0 & \sqrt{3} & 0 \\ \hline
 -1 & 0 & 0 & 0 & 0 & 0 & 0 & 0 \\
 \sqrt{3} & 0 & 0 & 0 & 0 & 0 & 0 & 0 \\
 0 & 0 & 0 & 0 & 0 & 0 & 0 & 0 \\
 0 & 0 & \sqrt{3} & 0 & 0 & 0 & 0 & 0 \\
 0 & \sqrt{3} & 0 & 0 & 0 & 0 & 0 & 0 \\
\end{array}
\right),
Q_{y}=\frac{1}{\sqrt{15}}
\left(
\begin{array}{ccc|ccccc}
 0 & 0 & 0 & 0 & 0 & 0 & 0 & \sqrt{3} \\
 0 & 0 & 0 & -1 & -\sqrt{3} & 0 & 0 & 0 \\
 0 & 0 & 0 & 0 & 0 & \sqrt{3} & 0 & 0 \\ \hline
 0 & -1 & 0 & 0 & 0 & 0 & 0 & 0 \\
 0 & -\sqrt{3} & 0 & 0 & 0 & 0 & 0 & 0 \\
 0 & 0 & \sqrt{3} & 0 & 0 & 0 & 0 & 0 \\
 0 & 0 & 0 & 0 & 0 & 0 & 0 & 0 \\
 \sqrt{3} & 0 & 0 & 0 & 0 & 0 & 0 & 0 \\
\end{array}
\right),
\cr&
Q_{z}=\frac{1}{\sqrt{15}}
\left(
\begin{array}{ccc|ccccc}
 0 & 0 & 0 & 0 & 0 & 0 & \sqrt{3} & 0 \\
 0 & 0 & 0 & 0 & 0 & \sqrt{3} & 0 & 0 \\
 0 & 0 & 0 & 2 & 0 & 0 & 0 & 0 \\ \hline
 0 & 0 & 2 & 0 & 0 & 0 & 0 & 0 \\
 0 & 0 & 0 & 0 & 0 & 0 & 0 & 0 \\
 0 & \sqrt{3} & 0 & 0 & 0 & 0 & 0 & 0 \\
 \sqrt{3} & 0 & 0 & 0 & 0 & 0 & 0 & 0 \\
 0 & 0 & 0 & 0 & 0 & 0 & 0 & 0 \\
\end{array}
\right),
\\&
T_{x}=\frac{2}{3\sqrt{15}}
\left(
\begin{array}{ccc|ccccc}
 0 & 0 & 0 & -i & i \sqrt{3} & 0 & 0 & 0 \\
 0 & 0 & 0 & 0 & 0 & 0 & 0 & i \sqrt{3} \\
 0 & 0 & 0 & 0 & 0 & 0 & i \sqrt{3} & 0 \\ \hline
 i & 0 & 0 & 0 & 0 & 0 & 0 & 0 \\
 -i \sqrt{3} & 0 & 0 & 0 & 0 & 0 & 0 & 0 \\
 0 & 0 & 0 & 0 & 0 & 0 & 0 & 0 \\
 0 & 0 & -i \sqrt{3} & 0 & 0 & 0 & 0 & 0 \\
 0 & -i \sqrt{3} & 0 & 0 & 0 & 0 & 0 & 0 \\
\end{array}
\right),
\cr&
T_{y}=\frac{2}{3\sqrt{15}}
\left(
\begin{array}{ccc|ccccc}
 0 & 0 & 0 & 0 & 0 & 0 & 0 & i \sqrt{3} \\
 0 & 0 & 0 & -i & -i \sqrt{3} & 0 & 0 & 0 \\
 0 & 0 & 0 & 0 & 0 & i \sqrt{3} & 0 & 0 \\ \hline
 0 & i & 0 & 0 & 0 & 0 & 0 & 0 \\
 0 & i \sqrt{3} & 0 & 0 & 0 & 0 & 0 & 0 \\
 0 & 0 & -i \sqrt{3} & 0 & 0 & 0 & 0 & 0 \\
 0 & 0 & 0 & 0 & 0 & 0 & 0 & 0 \\
 -i \sqrt{3} & 0 & 0 & 0 & 0 & 0 & 0 & 0 \\
\end{array}
\right),
\cr &
T_{z}=\frac{2}{3\sqrt{15}}
\left(
\begin{array}{ccc|ccccc}
 0 & 0 & 0 & 0 & 0 & 0 & i \sqrt{3} & 0 \\
 0 & 0 & 0 & 0 & 0 & i \sqrt{3} & 0 & 0 \\
 0 & 0 & 0 & 2 i & 0 & 0 & 0 & 0 \\ \hline
 0 & 0 & -2 i & 0 & 0 & 0 & 0 & 0 \\
 0 & 0 & 0 & 0 & 0 & 0 & 0 & 0 \\
 0 & -i \sqrt{3} & 0 & 0 & 0 & 0 & 0 & 0 \\
 -i \sqrt{3} & 0 & 0 & 0 & 0 & 0 & 0 & 0 \\
 0 & 0 & 0 & 0 & 0 & 0 & 0 & 0 \\
\end{array}
\right),
\end{align}
\begin{align}
&
M_{x}=
\left(
\begin{array}{ccc|ccccc}
 0 & 0 & 0 & 0 & 0 & 0 & 0 & 0 \\
 0 & 0 & -i & 0 & 0 & 0 & 0 & 0 \\
 0 & i & 0 & 0 & 0 & 0 & 0 & 0 \\ \hline
 0 & 0 & 0 & 0 & 0 & i \sqrt{3} & 0 & 0 \\
 0 & 0 & 0 & 0 & 0 & i & 0 & 0 \\
 0 & 0 & 0 & -i \sqrt{3} & -i & 0 & 0 & 0 \\
 0 & 0 & 0 & 0 & 0 & 0 & 0 & i \\
 0 & 0 & 0 & 0 & 0 & 0 & -i & 0 \\
\end{array}
\right),
M_{y}=
\left(
\begin{array}{ccc|ccccc}
 0 & 0 & i & 0 & 0 & 0 & 0 & 0 \\
 0 & 0 & 0 & 0 & 0 & 0 & 0 & 0 \\
 -i & 0 & 0 & 0 & 0 & 0 & 0 & 0 \\ \hline
 0 & 0 & 0 & 0 & 0 & 0 & -i \sqrt{3} & 0 \\
 0 & 0 & 0 & 0 & 0 & 0 & i & 0 \\
 0 & 0 & 0 & 0 & 0 & 0 & 0 & -i \\
 0 & 0 & 0 & i \sqrt{3} & -i & 0 & 0 & 0 \\
 0 & 0 & 0 & 0 & 0 & i & 0 & 0 \\
\end{array}
\right),
\cr&
M_{z}=
\left(
\begin{array}{ccc|ccccc}
 0 & -i & 0 & 0 & 0 & 0 & 0 & 0 \\
 i & 0 & 0 & 0 & 0 & 0 & 0 & 0 \\
 0 & 0 & 0 & 0 & 0 & 0 & 0 & 0 \\ \hline
 0 & 0 & 0 & 0 & 0 & 0 & 0 & 0 \\
 0 & 0 & 0 & 0 & 0 & 0 & 0 & -2 i \\
 0 & 0 & 0 & 0 & 0 & 0 & i & 0 \\
 0 & 0 & 0 & 0 & 0 & -i & 0 & 0 \\
 0 & 0 & 0 & 0 & 2 i & 0 & 0 & 0 \\
\end{array}
\right),
\\&
Q_{u}=\frac{1}{35}
\left(
\begin{array}{ccc|ccccc}
 -7 & 0 & 0 & 0 & 0 & 0 & 0 & 0 \\
 0 & -7 & 0 & 0 & 0 & 0 & 0 & 0 \\
 0 & 0 & 14 & 0 & 0 & 0 & 0 & 0 \\ \hline
 0 & 0 & 0 & 10 & 0 & 0 & 0 & 0 \\
 0 & 0 & 0 & 0 & -10 & 0 & 0 & 0 \\
 0 & 0 & 0 & 0 & 0 & 5 & 0 & 0 \\
 0 & 0 & 0 & 0 & 0 & 0 & 5 & 0 \\
 0 & 0 & 0 & 0 & 0 & 0 & 0 & -10 \\
\end{array}
\right),
Q_{v}=\frac{1}{35}
\left(
\begin{array}{ccc|ccccc}
 7 \sqrt{3} & 0 & 0 & 0 & 0 & 0 & 0 & 0 \\
 0 & -7 \sqrt{3} & 0 & 0 & 0 & 0 & 0 & 0 \\
 0 & 0 & 0 & 0 & 0 & 0 & 0 & 0 \\ \hline
 0 & 0 & 0 & 0 & -10 & 0 & 0 & 0 \\
 0 & 0 & 0 & -10 & 0 & 0 & 0 & 0 \\
 0 & 0 & 0 & 0 & 0 & -5 \sqrt{3} & 0 & 0 \\
 0 & 0 & 0 & 0 & 0 & 0 & 5 \sqrt{3} & 0 \\
 0 & 0 & 0 & 0 & 0 & 0 & 0 & 0 \\
\end{array}
\right), \nonumber
\\ &
Q_{yz}=\frac{1}{35}
\left(
\begin{array}{ccc|ccccc}
 0 & 0 & 0 & 0 & 0 & 0 & 0 & 0 \\
 0 & 0 & 7 \sqrt{3} & 0 & 0 & 0 & 0 & 0 \\
 0 & 7 \sqrt{3} & 0 & 0 & 0 & 0 & 0 & 0 \\ \hline
 0 & 0 & 0 & 0 & 0 & 5 & 0 & 0 \\
 0 & 0 & 0 & 0 & 0 & -5 \sqrt{3} & 0 & 0 \\
 0 & 0 & 0 & 5 & -5 \sqrt{3} & 0 & 0 & 0 \\
 0 & 0 & 0 & 0 & 0 & 0 & 0 & 5 \sqrt{3} \\
 0 & 0 & 0 & 0 & 0 & 0 & 5 \sqrt{3} & 0 \\
\end{array}
\right),
Q_{zx}=\frac{1}{35}
\left(
\begin{array}{ccc|ccccc}
 0 & 0 & 7 \sqrt{3} & 0 & 0 & 0 & 0 & 0 \\
 0 & 0 & 0 & 0 & 0 & 0 & 0 & 0 \\
 7 \sqrt{3} & 0 & 0 & 0 & 0 & 0 & 0 & 0 \\ \hline
 0 & 0 & 0 & 0 & 0 & 0 & 5 & 0 \\
 0 & 0 & 0 & 0 & 0 & 0 & 5 \sqrt{3} & 0 \\
 0 & 0 & 0 & 0 & 0 & 0 & 0 & 5 \sqrt{3} \\
 0 & 0 & 0 & 5 & 5 \sqrt{3} & 0 & 0 & 0 \\
 0 & 0 & 0 & 0 & 0 & 5 \sqrt{3} & 0 & 0 \\
\end{array}
\right),
\cr&
Q_{xy}=\frac{1}{35}
\left(
\begin{array}{ccc|ccccc}
 0 & 7 \sqrt{3} & 0 & 0 & 0 & 0 & 0 & 0 \\
 7 \sqrt{3} & 0 & 0 & 0 & 0 & 0 & 0 & 0 \\
 0 & 0 & 0 & 0 & 0 & 0 & 0 & 0 \\ \hline
 0 & 0 & 0 & 0 & 0 & 0 & 0 & -10 \\
 0 & 0 & 0 & 0 & 0 & 0 & 0 & 0 \\
 0 & 0 & 0 & 0 & 0 & 0 & 5 \sqrt{3} & 0 \\
 0 & 0 & 0 & 0 & 0 & 5 \sqrt{3} & 0 & 0 \\
 0 & 0 & 0 & -10 & 0 & 0 & 0 & 0 \\
\end{array}
\right),
\end{align}
\begin{align}
&
M_{u}=\frac{2}{\sqrt{5}}
\left(
\begin{array}{ccc|ccccc}
 0 & 0 & 0 & 0 & 0 & -i & 0 & 0 \\
 0 & 0 & 0 & 0 & 0 & 0 & i & 0 \\
 0 & 0 & 0 & 0 & 0 & 0 & 0 & 0 \\ \hline
 0 & 0 & 0 & 0 & 0 & 0 & 0 & 0 \\
 0 & 0 & 0 & 0 & 0 & 0 & 0 & 0 \\
 i & 0 & 0 & 0 & 0 & 0 & 0 & 0 \\
 0 & -i & 0 & 0 & 0 & 0 & 0 & 0 \\
 0 & 0 & 0 & 0 & 0 & 0 & 0 & 0 \\
\end{array}
\right),
M_{v}=\frac{2}{\sqrt{15}}
\left(
\begin{array}{ccc|ccccc}
 0 & 0 & 0 & 0 & 0 & -i & 0 & 0 \\
 0 & 0 & 0 & 0 & 0 & 0 & -i & 0 \\
 0 & 0 & 0 & 0 & 0 & 0 & 0 & 2 i \\ \hline
 0 & 0 & 0 & 0 & 0 & 0 & 0 & 0 \\
 0 & 0 & 0 & 0 & 0 & 0 & 0 & 0 \\
 i & 0 & 0 & 0 & 0 & 0 & 0 & 0 \\
 0 & i & 0 & 0 & 0 & 0 & 0 & 0 \\
 0 & 0 & -2 i & 0 & 0 & 0 & 0 & 0 \\
\end{array}
\right),
\cr&
M_{yz}=\frac{2}{\sqrt{15}}
\left(
\begin{array}{ccc|ccccc}
 0 & 0 & 0 & i \sqrt{3} & i & 0 & 0 & 0 \\
 0 & 0 & 0 & 0 & 0 & 0 & 0 & i \\
 0 & 0 & 0 & 0 & 0 & 0 & -i & 0 \\ \hline
 -i \sqrt{3} & 0 & 0 & 0 & 0 & 0 & 0 & 0 \\
 -i & 0 & 0 & 0 & 0 & 0 & 0 & 0 \\
 0 & 0 & 0 & 0 & 0 & 0 & 0 & 0 \\
 0 & 0 & i & 0 & 0 & 0 & 0 & 0 \\
 0 & -i & 0 & 0 & 0 & 0 & 0 & 0 \\
\end{array}
\right),
M_{zx}=\frac{2}{\sqrt{15}}
\left(
\begin{array}{ccc|ccccc}
 0 & 0 & 0 & 0 & 0 & 0 & 0 & -i \\
 0 & 0 & 0 & -i \sqrt{3} & i & 0 & 0 & 0 \\
 0 & 0 & 0 & 0 & 0 & i & 0 & 0 \\ \hline
 0 & i \sqrt{3} & 0 & 0 & 0 & 0 & 0 & 0 \\
 0 & -i & 0 & 0 & 0 & 0 & 0 & 0 \\
 0 & 0 & -i & 0 & 0 & 0 & 0 & 0 \\
 0 & 0 & 0 & 0 & 0 & 0 & 0 & 0 \\
 i & 0 & 0 & 0 & 0 & 0 & 0 & 0 \\
\end{array}
\right),
\cr&
M_{xy}=\frac{2}{\sqrt{15}}
\left(
\begin{array}{ccc|ccccc}
 0 & 0 & 0 & 0 & 0 & 0 & i & 0 \\
 0 & 0 & 0 & 0 & 0 & -i & 0 & 0 \\
 0 & 0 & 0 & 0 & -2 i & 0 & 0 & 0 \\ \hline
 0 & 0 & 0 & 0 & 0 & 0 & 0 & 0 \\
 0 & 0 & 2 i & 0 & 0 & 0 & 0 & 0 \\
 0 & i & 0 & 0 & 0 & 0 & 0 & 0 \\
 -i & 0 & 0 & 0 & 0 & 0 & 0 & 0 \\
 0 & 0 & 0 & 0 & 0 & 0 & 0 & 0 \\
\end{array}
\right),
\\&
G_{u}=\frac{2}{3\sqrt{5}}
\left(
\begin{array}{ccc|ccccc}
 0 & 0 & 0 & 0 & 0 & 1 & 0 & 0 \\
 0 & 0 & 0 & 0 & 0 & 0 & -1 & 0 \\
 0 & 0 & 0 & 0 & 0 & 0 & 0 & 0 \\ \hline
 0 & 0 & 0 & 0 & 0 & 0 & 0 & 0 \\
 0 & 0 & 0 & 0 & 0 & 0 & 0 & 0 \\
 1 & 0 & 0 & 0 & 0 & 0 & 0 & 0 \\
 0 & -1 & 0 & 0 & 0 & 0 & 0 & 0 \\
 0 & 0 & 0 & 0 & 0 & 0 & 0 & 0 \\
\end{array}
\right),
G_{v}=\frac{2}{3\sqrt{15}}
\left(
\begin{array}{ccc|ccccc}
 0 & 0 & 0 & 0 & 0 & 1 & 0 & 0 \\
 0 & 0 & 0 & 0 & 0 & 0 & 1 & 0 \\
 0 & 0 & 0 & 0 & 0 & 0 & 0 & -2 \\ \hline
 0 & 0 & 0 & 0 & 0 & 0 & 0 & 0 \\
 0 & 0 & 0 & 0 & 0 & 0 & 0 & 0 \\
 1 & 0 & 0 & 0 & 0 & 0 & 0 & 0 \\
 0 & 1 & 0 & 0 & 0 & 0 & 0 & 0 \\
 0 & 0 & -2 & 0 & 0 & 0 & 0 & 0 \\
\end{array}
\right),
\cr&
G_{yz}=\frac{2}{3\sqrt{15}}
\left(
\begin{array}{ccc|ccccc}
 0 & 0 & 0 & -\sqrt{3} & -1 & 0 & 0 & 0 \\
 0 & 0 & 0 & 0 & 0 & 0 & 0 & -1 \\
 0 & 0 & 0 & 0 & 0 & 0 & 1 & 0 \\ \hline
 -\sqrt{3} & 0 & 0 & 0 & 0 & 0 & 0 & 0 \\
 -1 & 0 & 0 & 0 & 0 & 0 & 0 & 0 \\
 0 & 0 & 0 & 0 & 0 & 0 & 0 & 0 \\
 0 & 0 & 1 & 0 & 0 & 0 & 0 & 0 \\
 0 & -1 & 0 & 0 & 0 & 0 & 0 & 0 \\
\end{array}
\right),
G_{zx}=\frac{2}{3\sqrt{15}}
\left(
\begin{array}{ccc|ccccc}
 0 & 0 & 0 & 0 & 0 & 0 & 0 & 1 \\
 0 & 0 & 0 & \sqrt{3} & -1 & 0 & 0 & 0 \\
 0 & 0 & 0 & 0 & 0 & -1 & 0 & 0 \\ \hline
 0 & \sqrt{3} & 0 & 0 & 0 & 0 & 0 & 0 \\
 0 & -1 & 0 & 0 & 0 & 0 & 0 & 0 \\
 0 & 0 & -1 & 0 & 0 & 0 & 0 & 0 \\
 0 & 0 & 0 & 0 & 0 & 0 & 0 & 0 \\
 1 & 0 & 0 & 0 & 0 & 0 & 0 & 0 \\
\end{array}
\right),
\cr&
G_{xy}=\frac{2}{3\sqrt{15}}
\left(
\begin{array}{ccc|ccccc}
 0 & 0 & 0 & 0 & 0 & 0 & -1 & 0 \\
 0 & 0 & 0 & 0 & 0 & 1 & 0 & 0 \\
 0 & 0 & 0 & 0 & 2 & 0 & 0 & 0 \\ \hline
 0 & 0 & 0 & 0 & 0 & 0 & 0 & 0 \\
 0 & 0 & 2 & 0 & 0 & 0 & 0 & 0 \\
 0 & 1 & 0 & 0 & 0 & 0 & 0 & 0 \\
 -1 & 0 & 0 & 0 & 0 & 0 & 0 & 0 \\
 0 & 0 & 0 & 0 & 0 & 0 & 0 & 0 \\
\end{array}
\right), 
\end{align}
\begin{align}
&
Q_{xyz}=\frac{\sqrt{3}}{7}
\left(
\begin{array}{ccc|ccccc}
 0 & 0 & 0 & 0 & 0 & 1 & 0 & 0 \\
 0 & 0 & 0 & 0 & 0 & 0 & 1 & 0 \\
 0 & 0 & 0 & 0 & 0 & 0 & 0 & 1 \\ \hline
 0 & 0 & 0 & 0 & 0 & 0 & 0 & 0 \\
 0 & 0 & 0 & 0 & 0 & 0 & 0 & 0 \\
 1 & 0 & 0 & 0 & 0 & 0 & 0 & 0 \\
 0 & 1 & 0 & 0 & 0 & 0 & 0 & 0 \\
 0 & 0 & 1 & 0 & 0 & 0 & 0 & 0 \\
\end{array}
\right),
Q_{x}^{\alpha}=\frac{3}{14\sqrt{5}}
\left(
\begin{array}{ccc|ccccc}
 0 & 0 & 0 & -\sqrt{3} & 3 & 0 & 0 & 0 \\
 0 & 0 & 0 & 0 & 0 & 0 & 0 & -2 \\
 0 & 0 & 0 & 0 & 0 & 0 & -2 & 0 \\ \hline
 -\sqrt{3} & 0 & 0 & 0 & 0 & 0 & 0 & 0 \\
 3 & 0 & 0 & 0 & 0 & 0 & 0 & 0 \\
 0 & 0 & 0 & 0 & 0 & 0 & 0 & 0 \\
 0 & 0 & -2 & 0 & 0 & 0 & 0 & 0 \\
 0 & -2 & 0 & 0 & 0 & 0 & 0 & 0 \\
\end{array}
\right),
\cr&
Q_{y}^{\alpha}=\frac{3}{14\sqrt{5}}
\left(
\begin{array}{ccc|ccccc}
 0 & 0 & 0 & 0 & 0 & 0 & 0 & -2 \\
 0 & 0 & 0 & -\sqrt{3} & -3 & 0 & 0 & 0 \\
 0 & 0 & 0 & 0 & 0 & -2 & 0 & 0 \\ \hline
 0 & -\sqrt{3} & 0 & 0 & 0 & 0 & 0 & 0 \\
 0 & -3 & 0 & 0 & 0 & 0 & 0 & 0 \\
 0 & 0 & -2 & 0 & 0 & 0 & 0 & 0 \\
 0 & 0 & 0 & 0 & 0 & 0 & 0 & 0 \\
 -2 & 0 & 0 & 0 & 0 & 0 & 0 & 0 \\
\end{array}
\right),
Q_{z}^{\alpha}=\frac{3}{7\sqrt{5}}
\left(
\begin{array}{ccc|ccccc}
 0 & 0 & 0 & 0 & 0 & 0 & -1 & 0 \\
 0 & 0 & 0 & 0 & 0 & -1 & 0 & 0 \\
 0 & 0 & 0 & \sqrt{3} & 0 & 0 & 0 & 0 \\ \hline
 0 & 0 & \sqrt{3} & 0 & 0 & 0 & 0 & 0 \\
 0 & 0 & 0 & 0 & 0 & 0 & 0 & 0 \\
 0 & -1 & 0 & 0 & 0 & 0 & 0 & 0 \\
 -1 & 0 & 0 & 0 & 0 & 0 & 0 & 0 \\
 0 & 0 & 0 & 0 & 0 & 0 & 0 & 0 \\
\end{array}
\right),
\cr&
Q_{x}^{\beta}=\frac{\sqrt{3}}{14}
\left(
\begin{array}{ccc|ccccc}
 0 & 0 & 0 & -\sqrt{3} & -1 & 0 & 0 & 0 \\
 0 & 0 & 0 & 0 & 0 & 0 & 0 & 2 \\
 0 & 0 & 0 & 0 & 0 & 0 & -2 & 0 \\ \hline
 -\sqrt{3} & 0 & 0 & 0 & 0 & 0 & 0 & 0 \\
 -1 & 0 & 0 & 0 & 0 & 0 & 0 & 0 \\
 0 & 0 & 0 & 0 & 0 & 0 & 0 & 0 \\
 0 & 0 & -2 & 0 & 0 & 0 & 0 & 0 \\
 0 & 2 & 0 & 0 & 0 & 0 & 0 & 0 \\
\end{array}
\right),
Q_{y}^{\beta}=\frac{\sqrt{3}}{14}
\left(
\begin{array}{ccc|ccccc}
 0 & 0 & 0 & 0 & 0 & 0 & 0 & -2 \\
 0 & 0 & 0 & \sqrt{3} & -1 & 0 & 0 & 0 \\
 0 & 0 & 0 & 0 & 0 & 2 & 0 & 0 \\ \hline
 0 & \sqrt{3} & 0 & 0 & 0 & 0 & 0 & 0 \\
 0 & -1 & 0 & 0 & 0 & 0 & 0 & 0 \\
 0 & 0 & 2 & 0 & 0 & 0 & 0 & 0 \\
 0 & 0 & 0 & 0 & 0 & 0 & 0 & 0 \\
 -2 & 0 & 0 & 0 & 0 & 0 & 0 & 0 \\
\end{array}
\right),
\cr&
Q_{z}^{\beta}=\frac{\sqrt{3}}{7}
\left(
\begin{array}{ccc|ccccc}
 0 & 0 & 0 & 0 & 0 & 0 & 1 & 0 \\
 0 & 0 & 0 & 0 & 0 & -1 & 0 & 0 \\
 0 & 0 & 0 & 0 & 1 & 0 & 0 & 0 \\ \hline
 0 & 0 & 0 & 0 & 0 & 0 & 0 & 0 \\
 0 & 0 & 1 & 0 & 0 & 0 & 0 & 0 \\
 0 & -1 & 0 & 0 & 0 & 0 & 0 & 0 \\
 1 & 0 & 0 & 0 & 0 & 0 & 0 & 0 \\
 0 & 0 & 0 & 0 & 0 & 0 & 0 & 0 \\
\end{array}
\right),
\end{align}
\begin{align}
&
T_{xyz}=\frac{\sqrt{3}}{35}
\left(
\begin{array}{ccc|ccccc}
 0 & 0 & 0 & 0 & 0 & i & 0 & 0 \\
 0 & 0 & 0 & 0 & 0 & 0 & i & 0 \\
 0 & 0 & 0 & 0 & 0 & 0 & 0 & i \\ \hline
 0 & 0 & 0 & 0 & 0 & 0 & 0 & 0 \\
 0 & 0 & 0 & 0 & 0 & 0 & 0 & 0 \\
-i & 0 & 0 & 0 & 0 & 0 & 0 & 0 \\
 0 &-i & 0 & 0 & 0 & 0 & 0 & 0 \\
 0 & 0 &-i & 0 & 0 & 0 & 0 & 0 \\
\end{array}
\right),
T_{x}^{\alpha}=\frac{3}{70\sqrt{5}}
\left(
\begin{array}{ccc|ccccc}
 0 & 0 & 0 & -\sqrt{3} i & 3 i  & 0 & 0 & 0 \\
 0 & 0 & 0 & 0 & 0 & 0 & 0 & -2 i  \\
 0 & 0 & 0 & 0 & 0 & 0 & -2 i & 0 \\ \hline
 \sqrt{3} i & 0 & 0 & 0 & 0 & 0 & 0 & 0 \\
 -3 i  & 0 & 0 & 0 & 0 & 0 & 0 & 0 \\
 0 & 0 & 0 & 0 & 0 & 0 & 0 & 0 \\
 0 & 0 & 2 i & 0 & 0 & 0 & 0 & 0 \\
 0 & 2 i & 0 & 0 & 0 & 0 & 0 & 0 \\
\end{array}
\right),
\cr&
T_{y}^{\alpha}=\frac{3}{70\sqrt{5}}
\left(
\begin{array}{ccc|ccccc}
 0 & 0 & 0 & 0 & 0 & 0 & 0 & -2 i \\
 0 & 0 & 0 & -i \sqrt{3} & -3 i & 0 & 0 & 0 \\
 0 & 0 & 0 & 0 & 0 & -2 i & 0 & 0 \\ \hline
 0 & i \sqrt{3} & 0 & 0 & 0 & 0 & 0 & 0 \\
 0 & 3 i & 0 & 0 & 0 & 0 & 0 & 0 \\
 0 & 0 & 2 i & 0 & 0 & 0 & 0 & 0 \\
 0 & 0 & 0 & 0 & 0 & 0 & 0 & 0 \\
 2 i & 0 & 0 & 0 & 0 & 0 & 0 & 0 \\
\end{array}
\right),
T_{z}^{\alpha}=\frac{3}{35\sqrt{5}}
\left(
\begin{array}{ccc|ccccc}
 0 & 0 & 0 & 0 & 0 & 0 & - i & 0 \\
 0 & 0 & 0 & 0 & 0 & - i & 0 & 0 \\
 0 & 0 & 0 &  i \sqrt{3} & 0 & 0 & 0 & 0 \\ \hline
 0 & 0 & - i \sqrt{3} & 0 & 0 & 0 & 0 & 0 \\
 0 & 0 & 0 & 0 & 0 & 0 & 0 & 0 \\
 0 &  i & 0 & 0 & 0 & 0 & 0 & 0 \\
  i & 0 & 0 & 0 & 0 & 0 & 0 & 0 \\
 0 & 0 & 0 & 0 & 0 & 0 & 0 & 0 \\
\end{array}
\right),
\cr&
T_{x}^{\beta}=\frac{\sqrt{3}}{70}
\left(
\begin{array}{ccc|ccccc}
 0 & 0 & 0 & -i \sqrt{3} & -i & 0 & 0 & 0 \\
 0 & 0 & 0 & 0 & 0 & 0 & 0 & 2 i \\
 0 & 0 & 0 & 0 & 0 & 0 & -2 i & 0 \\ \hline
 i \sqrt{3} & 0 & 0 & 0 & 0 & 0 & 0 & 0 \\
 i & 0 & 0 & 0 & 0 & 0 & 0 & 0 \\
 0 & 0 & 0 & 0 & 0 & 0 & 0 & 0 \\
 0 & 0 & 2 i & 0 & 0 & 0 & 0 & 0 \\
 0 & -2 i & 0 & 0 & 0 & 0 & 0 & 0 \\
\end{array}
\right),
T_{y}^{\beta}=\frac{\sqrt{3}}{70}
\left(
\begin{array}{ccc|ccccc}
 0 & 0 & 0 & 0 & 0 & 0 & 0 & -2 i \\
 0 & 0 & 0 & i \sqrt{3} & -i & 0 & 0 & 0 \\
 0 & 0 & 0 & 0 & 0 & 2 i & 0 & 0 \\ \hline
 0 & -i \sqrt{3} & 0 & 0 & 0 & 0 & 0 & 0 \\
 0 & i & 0 & 0 & 0 & 0 & 0 & 0 \\
 0 & 0 & -2 i & 0 & 0 & 0 & 0 & 0 \\
 0 & 0 & 0 & 0 & 0 & 0 & 0 & 0 \\
 2 i & 0 & 0 & 0 & 0 & 0 & 0 & 0 \\
\end{array}
\right),
\cr&
T_{z}^{\beta}=\frac{\sqrt{3}}{35}
\left(
\begin{array}{ccc|ccccc}
 0 & 0 & 0 & 0 & 0 & 0 & i & 0 \\
 0 & 0 & 0 & 0 & 0 & -i & 0 & 0 \\
 0 & 0 & 0 & 0 & i & 0 & 0 & 0 \\ \hline
 0 & 0 & 0 & 0 & 0 & 0 & 0 & 0 \\
 0 & 0 & -i & 0 & 0 & 0 & 0 & 0 \\
 0 & i & 0 & 0 & 0 & 0 & 0 & 0 \\
 -i & 0 & 0 & 0 & 0 & 0 & 0 & 0 \\
 0 & 0 & 0 & 0 & 0 & 0 & 0 & 0 \\
\end{array}
\right). 
\end{align}
\begin{align}
&
M_{xyz}=\frac{\sqrt{45}}{7}
\left(
\begin{array}{ccc|ccccc}
 0 & 0 & 0 & 0 & 0 & 0 & 0 & 0 \\
 0 & 0 & 0 & 0 & 0 & 0 & 0 & 0 \\
 0 & 0 & 0 & 0 & 0 & 0 & 0 & 0 \\ \hline
 0 & 0 & 0 & 0 & -i & 0 & 0 & 0 \\
 0 & 0 & 0 & i & 0 & 0 & 0 & 0 \\
 0 & 0 & 0 & 0 & 0 & 0 & 0 & 0 \\
 0 & 0 & 0 & 0 & 0 & 0 & 0 & 0 \\
 0 & 0 & 0 & 0 & 0 & 0 & 0 & 0 \\
\end{array}
\right),
M_{x}^{\alpha}=\frac{3}{14}
\left(
\begin{array}{ccc|ccccc}
 0 & 0 & 0 & 0 & 0 & 0 & 0 & 0 \\
 0 & 0 & 0 & 0 & 0 & 0 & 0 & 0 \\
 0 & 0 & 0 & 0 & 0 & 0 & 0 & 0 \\ \hline
 0 & 0 & 0 & 0 & 0 & -i \sqrt{3} & 0 & 0 \\
 0 & 0 & 0 & 0 & 0 & -i & 0 & 0 \\
 0 & 0 & 0 & i \sqrt{3} & i & 0 & 0 & 0 \\
 0 & 0 & 0 & 0 & 0 & 0 & 0 & 4 i \\
 0 & 0 & 0 & 0 & 0 & 0 & -4 i & 0 \\
\end{array}
\right),
\cr&
M_{y}^{\alpha}=\frac{3}{14}
\left(
\begin{array}{ccc|ccccc}
 0 & 0 & 0 & 0 & 0 & 0 & 0 & 0 \\
 0 & 0 & 0 & 0 & 0 & 0 & 0 & 0 \\
 0 & 0 & 0 & 0 & 0 & 0 & 0 & 0 \\ \hline
 0 & 0 & 0 & 0 & 0 & 0 & i \sqrt{3} & 0 \\
 0 & 0 & 0 & 0 & 0 & 0 & -i & 0 \\
 0 & 0 & 0 & 0 & 0 & 0 & 0 & -4 i \\
 0 & 0 & 0 & -i \sqrt{3} & i & 0 & 0 & 0 \\
 0 & 0 & 0 & 0 & 0 & 4 i & 0 & 0 \\
\end{array}
\right),
M_{z}^{\alpha}=\frac{3}{7}
\left(
\begin{array}{ccc|ccccc}
 0 & 0 & 0 & 0 & 0 & 0 & 0 & 0 \\
 0 & 0 & 0 & 0 & 0 & 0 & 0 & 0 \\
 0 & 0 & 0 & 0 & 0 & 0 & 0 & 0 \\ \hline
 0 & 0 & 0 & 0 & 0 & 0 & 0 & 0 \\
 0 & 0 & 0 & 0 & 0 & 0 & 0 & i \\
 0 & 0 & 0 & 0 & 0 & 0 & 2 i & 0 \\
 0 & 0 & 0 & 0 & 0 & -2 i & 0 & 0 \\
 0 & 0 & 0 & 0 & -i & 0 & 0 & 0 \\
\end{array}
\right),
\cr&
M_{x}^{\beta}=\frac{3\sqrt{5}}{14}
\left(
\begin{array}{ccc|ccccc}
 0 & 0 & 0 & 0 & 0 & 0 & 0 & 0 \\
 0 & 0 & 0 & 0 & 0 & 0 & 0 & 0 \\
 0 & 0 & 0 & 0 & 0 & 0 & 0 & 0 \\ \hline
 0 & 0 & 0 & 0 & 0 & -i & 0 & 0 \\
 0 & 0 & 0 & 0 & 0 & i \sqrt{3} & 0 & 0 \\
 0 & 0 & 0 & i & -i \sqrt{3} & 0 & 0 & 0 \\
 0 & 0 & 0 & 0 & 0 & 0 & 0 & 0 \\
 0 & 0 & 0 & 0 & 0 & 0 & 0 & 0 \\
\end{array}
\right),
M_{y}^{\beta}=\frac{3\sqrt{5}}{14}
\left(
\begin{array}{ccc|ccccc}
 0 & 0 & 0 & 0 & 0 & 0 & 0 & 0 \\
 0 & 0 & 0 & 0 & 0 & 0 & 0 & 0 \\
 0 & 0 & 0 & 0 & 0 & 0 & 0 & 0 \\ \hline
 0 & 0 & 0 & 0 & 0 & 0 & -i & 0 \\
 0 & 0 & 0 & 0 & 0 & 0 & -i \sqrt{3} & 0 \\
 0 & 0 & 0 & 0 & 0 & 0 & 0 & 0 \\
 0 & 0 & 0 & i & i \sqrt{3} & 0 & 0 & 0 \\
 0 & 0 & 0 & 0 & 0 & 0 & 0 & 0 \\
\end{array}
\right),
\cr&
M_{z}^{\beta}=\frac{3\sqrt{5}}{7}
\left(
\begin{array}{ccc|ccccc}
 0 & 0 & 0 & 0 & 0 & 0 & 0 & 0 \\
 0 & 0 & 0 & 0 & 0 & 0 & 0 & 0 \\
 0 & 0 & 0 & 0 & 0 & 0 & 0 & 0 \\ \hline
 0 & 0 & 0 & 0 & 0 & 0 & 0 & i \\
 0 & 0 & 0 & 0 & 0 & 0 & 0 & 0 \\
 0 & 0 & 0 & 0 & 0 & 0 & 0 & 0 \\
 0 & 0 & 0 & 0 & 0 & 0 & 0 & 0 \\
 0 & 0 & 0 & -i & 0 & 0 & 0 & 0 \\
\end{array}
\right). 
\end{align}

\subsection{$d$-$f$ hybridized-orbital system}
Matrix elements of each multipole activated in the $d$-$f$ hybridized-orbital system are given below. 
The basis function is taken as $(\phi_{u},\phi_{v},\phi_{yz},\phi_{zx},\phi_{xy},\phi_{xyz},\phi_{x}^{\alpha},\phi_{y}^{\alpha},\phi_{z}^{\alpha},\phi_{x}^{\beta},\phi_{y}^{\beta},\phi_{z}^{\beta})$. 

\begin{align}
&
Q_{x}=\frac{1}{2\sqrt{35}}
\left(
\begin{array}{ccccc|ccccccc}
 0 & 0 & 0 & 0 & 0 & 0 & -3 & 0 & 0 & -\sqrt{15} & 0 & 0 \\
 0 & 0 & 0 & 0 & 0 & 0 & 3 \sqrt{3} & 0 & 0 & -\sqrt{5} & 0 & 0 \\
 0 & 0 & 0 & 0 & 0 & 2 \sqrt{5} & 0 & 0 & 0 & 0 & 0 & 0 \\
 0 & 0 & 0 & 0 & 0 & 0 & 0 & 0 & -2 \sqrt{3} & 0 & 0 & 2 \sqrt{5} \\
 0 & 0 & 0 & 0 & 0 & 0 & 0 & -2 \sqrt{3} & 0 & 0 & -2 \sqrt{5} & 0 \\ \hline
 0 & 0 & 2 \sqrt{5} & 0 & 0 & 0 & 0 & 0 & 0 & 0 & 0 & 0 \\
 -3 & 3 \sqrt{3} & 0 & 0 & 0 & 0 & 0 & 0 & 0 & 0 & 0 & 0 \\
 0 & 0 & 0 & 0 & -2 \sqrt{3} & 0 & 0 & 0 & 0 & 0 & 0 & 0 \\
 0 & 0 & 0 & -2 \sqrt{3} & 0 & 0 & 0 & 0 & 0 & 0 & 0 & 0 \\
 -\sqrt{15} & -\sqrt{5} & 0 & 0 & 0 & 0 & 0 & 0 & 0 & 0 & 0 & 0 \\
 0 & 0 & 0 & 0 & -2 \sqrt{5} & 0 & 0 & 0 & 0 & 0 & 0 & 0 \\
 0 & 0 & 0 & 2 \sqrt{5} & 0 & 0 & 0 & 0 & 0 & 0 & 0 & 0 \\
\end{array}
\right), 
\cr&
Q_{y}=\frac{1}{2\sqrt{35}}
\left(
\begin{array}{ccccc|ccccccc}
 0 & 0 & 0 & 0 & 0 & 0 & 0 & -3 & 0 & 0 & \sqrt{15} & 0 \\
 0 & 0 & 0 & 0 & 0 & 0 & 0 & -3 \sqrt{3} & 0 & 0 & -\sqrt{5} & 0 \\
 0 & 0 & 0 & 0 & 0 & 0 & 0 & 0 & -2 \sqrt{3} & 0 & 0 & -2 \sqrt{5} \\
 0 & 0 & 0 & 0 & 0 & 2 \sqrt{5} & 0 & 0 & 0 & 0 & 0 & 0 \\
 0 & 0 & 0 & 0 & 0 & 0 & -2 \sqrt{3} & 0 & 0 & 2 \sqrt{5} & 0 & 0 \\ \hline
 0 & 0 & 0 & 2 \sqrt{5} & 0 & 0 & 0 & 0 & 0 & 0 & 0 & 0 \\
 0 & 0 & 0 & 0 & -2 \sqrt{3} & 0 & 0 & 0 & 0 & 0 & 0 & 0 \\
 -3 & -3 \sqrt{3} & 0 & 0 & 0 & 0 & 0 & 0 & 0 & 0 & 0 & 0 \\
 0 & 0 & -2 \sqrt{3} & 0 & 0 & 0 & 0 & 0 & 0 & 0 & 0 & 0 \\
 0 & 0 & 0 & 0 & 2 \sqrt{5} & 0 & 0 & 0 & 0 & 0 & 0 & 0 \\
 \sqrt{15} & -\sqrt{5} & 0 & 0 & 0 & 0 & 0 & 0 & 0 & 0 & 0 & 0 \\
 0 & 0 & -2 \sqrt{5} & 0 & 0 & 0 & 0 & 0 & 0 & 0 & 0 & 0 \\
\end{array}
\right), 
\cr&
Q_{z}=\frac{1}{\sqrt{35}}
\left(
\begin{array}{ccccc|ccccccc}
 0 & 0 & 0 & 0 & 0 & 0 & 0 & 0 & 3 & 0 & 0 & 0 \\
 0 & 0 & 0 & 0 & 0 & 0 & 0 & 0 & 0 & 0 & 0 & \sqrt{5} \\
 0 & 0 & 0 & 0 & 0 & 0 & 0 & -\sqrt{3} & 0 & 0 & \sqrt{5} & 0 \\
 0 & 0 & 0 & 0 & 0 & 0 & -\sqrt{3} & 0 & 0 & -\sqrt{5} & 0 & 0 \\
 0 & 0 & 0 & 0 & 0 & \sqrt{5} & 0 & 0 & 0 & 0 & 0 & 0 \\ \hline
 0 & 0 & 0 & 0 & \sqrt{5} & 0 & 0 & 0 & 0 & 0 & 0 & 0 \\
 0 & 0 & 0 & -\sqrt{3} & 0 & 0 & 0 & 0 & 0 & 0 & 0 & 0 \\
 0 & 0 & -\sqrt{3} & 0 & 0 & 0 & 0 & 0 & 0 & 0 & 0 & 0 \\
 3 & 0 & 0 & 0 & 0 & 0 & 0 & 0 & 0 & 0 & 0 & 0 \\
 0 & 0 & 0 & -\sqrt{5} & 0 & 0 & 0 & 0 & 0 & 0 & 0 & 0 \\
 0 & 0 & \sqrt{5} & 0 & 0 & 0 & 0 & 0 & 0 & 0 & 0 & 0 \\
 0 & \sqrt{5} & 0 & 0 & 0 & 0 & 0 & 0 & 0 & 0 & 0 & 0 \\
\end{array}
\right),  
\end{align}

\begin{align}
&
T_{x}=\frac{1}{2\sqrt{35}}
\left(
\begin{array}{ccccc|ccccccc}
 0 & 0 & 0 & 0 & 0 & 0 & -3 i & 0 & 0 & -i \sqrt{15} & 0 & 0 \\
 0 & 0 & 0 & 0 & 0 & 0 & 3 i \sqrt{3} & 0 & 0 & -i \sqrt{5} & 0 & 0 \\
 0 & 0 & 0 & 0 & 0 & 2 i \sqrt{5} & 0 & 0 & 0 & 0 & 0 & 0 \\
 0 & 0 & 0 & 0 & 0 & 0 & 0 & 0 & -2 i \sqrt{3} & 0 & 0 & 2 i \sqrt{5} \\
 0 & 0 & 0 & 0 & 0 & 0 & 0 & -2 i \sqrt{3} & 0 & 0 & -2 i \sqrt{5} & 0 \\ \hline
 0 & 0 & -2 i \sqrt{5} & 0 & 0 & 0 & 0 & 0 & 0 & 0 & 0 & 0 \\
 3 i & -3 i \sqrt{3} & 0 & 0 & 0 & 0 & 0 & 0 & 0 & 0 & 0 & 0 \\
 0 & 0 & 0 & 0 & 2 i \sqrt{3} & 0 & 0 & 0 & 0 & 0 & 0 & 0 \\
 0 & 0 & 0 & 2 i \sqrt{3} & 0 & 0 & 0 & 0 & 0 & 0 & 0 & 0 \\
 i \sqrt{15} & i \sqrt{5} & 0 & 0 & 0 & 0 & 0 & 0 & 0 & 0 & 0 & 0 \\
 0 & 0 & 0 & 0 & 2 i \sqrt{5} & 0 & 0 & 0 & 0 & 0 & 0 & 0 \\
 0 & 0 & 0 & -2 i \sqrt{5} & 0 & 0 & 0 & 0 & 0 & 0 & 0 & 0 \\
\end{array}
\right), 
\cr&
T_{y}=\frac{1}{2\sqrt{35}}
\left(
\begin{array}{ccccc|ccccccc}
 0 & 0 & 0 & 0 & 0 & 0 & 0 & -3 i & 0 & 0 & i \sqrt{15} & 0 \\
 0 & 0 & 0 & 0 & 0 & 0 & 0 & -3 i \sqrt{3} & 0 & 0 & -i \sqrt{5} & 0 \\
 0 & 0 & 0 & 0 & 0 & 0 & 0 & 0 & -2 i \sqrt{3} & 0 & 0 & -2 i \sqrt{5} \\
 0 & 0 & 0 & 0 & 0 & 2 i \sqrt{5} & 0 & 0 & 0 & 0 & 0 & 0 \\
 0 & 0 & 0 & 0 & 0 & 0 & -2 i \sqrt{3} & 0 & 0 & 2 i \sqrt{5} & 0 & 0 \\ \hline
 0 & 0 & 0 & -2 i \sqrt{5} & 0 & 0 & 0 & 0 & 0 & 0 & 0 & 0 \\
 0 & 0 & 0 & 0 & 2 i \sqrt{3} & 0 & 0 & 0 & 0 & 0 & 0 & 0 \\
 3 i & 3 i \sqrt{3} & 0 & 0 & 0 & 0 & 0 & 0 & 0 & 0 & 0 & 0 \\
 0 & 0 & 2 i \sqrt{3} & 0 & 0 & 0 & 0 & 0 & 0 & 0 & 0 & 0 \\
 0 & 0 & 0 & 0 & -2 i \sqrt{5} & 0 & 0 & 0 & 0 & 0 & 0 & 0 \\
 -i \sqrt{15} & i \sqrt{5} & 0 & 0 & 0 & 0 & 0 & 0 & 0 & 0 & 0 & 0 \\
 0 & 0 & 2 i \sqrt{5} & 0 & 0 & 0 & 0 & 0 & 0 & 0 & 0 & 0 \\
\end{array}
\right), 
\cr&
T_{z}=\frac{1}{\sqrt{35}}
\left(
\begin{array}{ccccc|ccccccc}
 0 & 0 & 0 & 0 & 0 & 0 & 0 & 0 & 3 i & 0 & 0 & 0 \\
 0 & 0 & 0 & 0 & 0 & 0 & 0 & 0 & 0 & 0 & 0 & i \sqrt{5} \\
 0 & 0 & 0 & 0 & 0 & 0 & 0 & -i \sqrt{3} & 0 & 0 & i \sqrt{5} & 0 \\
 0 & 0 & 0 & 0 & 0 & 0 & -i \sqrt{3} & 0 & 0 & -i \sqrt{5} & 0 & 0 \\
 0 & 0 & 0 & 0 & 0 & i \sqrt{5} & 0 & 0 & 0 & 0 & 0 & 0 \\ \hline
 0 & 0 & 0 & 0 & -i \sqrt{5} & 0 & 0 & 0 & 0 & 0 & 0 & 0 \\
 0 & 0 & 0 & i \sqrt{3} & 0 & 0 & 0 & 0 & 0 & 0 & 0 & 0 \\
 0 & 0 & i \sqrt{3} & 0 & 0 & 0 & 0 & 0 & 0 & 0 & 0 & 0 \\
 -3 i & 0 & 0 & 0 & 0 & 0 & 0 & 0 & 0 & 0 & 0 & 0 \\
 0 & 0 & 0 & i \sqrt{5} & 0 & 0 & 0 & 0 & 0 & 0 & 0 & 0 \\
 0 & 0 & -i \sqrt{5} & 0 & 0 & 0 & 0 & 0 & 0 & 0 & 0 & 0 \\
 0 & -i \sqrt{5} & 0 & 0 & 0 & 0 & 0 & 0 & 0 & 0 & 0 & 0 \\
\end{array}
\right), 
\end{align}

\begin{align}
&
M_{u}=\frac{2}{\sqrt{35}}
\scalebox{0.95}{$\displaystyle
\left(
\begin{array}{ccccc|ccccccc}
 0 & 0 & 0 & 0 & 0 & 0 & 0 & 0 & 0 & 0 & 0 & 0 \\
 0 & 0 & 0 & 0 & 0 & -2 i \sqrt{5} & 0 & 0 & 0 & 0 & 0 & 0 \\
 0 & 0 & 0 & 0 & 0 & 0 & -i \sqrt{3} & 0 & 0 & -i \sqrt{5} & 0 & 0 \\
 0 & 0 & 0 & 0 & 0 & 0 & 0 & i \sqrt{3} & 0 & 0 & -i \sqrt{5} & 0 \\
 0 & 0 & 0 & 0 & 0 & 0 & 0 & 0 & 0 & 0 & 0 & 2 i \sqrt{5} \\ \hline
 0 & 2 i \sqrt{5} & 0 & 0 & 0 & 0 & 0 & 0 & 0 & 0 & 0 & 0 \\
 0 & 0 & i \sqrt{3} & 0 & 0 & 0 & 0 & 0 & 0 & 0 & 0 & 0 \\
 0 & 0 & 0 & -i \sqrt{3} & 0 & 0 & 0 & 0 & 0 & 0 & 0 & 0 \\
 0 & 0 & 0 & 0 & 0 & 0 & 0 & 0 & 0 & 0 & 0 & 0 \\
 0 & 0 & i \sqrt{5} & 0 & 0 & 0 & 0 & 0 & 0 & 0 & 0 & 0 \\
 0 & 0 & 0 & i \sqrt{5} & 0 & 0 & 0 & 0 & 0 & 0 & 0 & 0 \\
 0 & 0 & 0 & 0 & -2 i \sqrt{5} & 0 & 0 & 0 & 0 & 0 & 0 & 0 \\
\end{array}
\right), 
$}
\cr&
M_{v}=\frac{2}{\sqrt{35}}
\scalebox{0.95}{$\displaystyle
\left(
\begin{array}{ccccc|ccccccc}
 0 & 0 & 0 & 0 & 0 & 2 i \sqrt{5} & 0 & 0 & 0 & 0 & 0 & 0 \\
 0 & 0 & 0 & 0 & 0 & 0 & 0 & 0 & 0 & 0 & 0 & 0 \\
 0 & 0 & 0 & 0 & 0 & 0 & -i & 0 & 0 & i \sqrt{15} & 0 & 0 \\
 0 & 0 & 0 & 0 & 0 & 0 & 0 & -i & 0 & 0 & -i \sqrt{15} & 0 \\
 0 & 0 & 0 & 0 & 0 & 0 & 0 & 0 & 2 i & 0 & 0 & 0 \\ \hline
 -2 i \sqrt{5} & 0 & 0 & 0 & 0 & 0 & 0 & 0 & 0 & 0 & 0 & 0 \\
 0 & 0 & i & 0 & 0 & 0 & 0 & 0 & 0 & 0 & 0 & 0 \\
 0 & 0 & 0 & i & 0 & 0 & 0 & 0 & 0 & 0 & 0 & 0 \\
 0 & 0 & 0 & 0 & -2 i & 0 & 0 & 0 & 0 & 0 & 0 & 0 \\
 0 & 0 & -i \sqrt{15} & 0 & 0 & 0 & 0 & 0 & 0 & 0 & 0 & 0 \\
 0 & 0 & 0 & i \sqrt{15} & 0 & 0 & 0 & 0 & 0 & 0 & 0 & 0 \\
 0 & 0 & 0 & 0 & 0 & 0 & 0 & 0 & 0 & 0 & 0 & 0 \\
\end{array}
\right), 
$}
\cr&
M_{yz}=\frac{2}{\sqrt{35}}
\scalebox{0.95}{$\displaystyle
\left(
\begin{array}{ccccc|ccccccc}
 0 & 0 & 0 & 0 & 0 & 0 & i \sqrt{3} & 0 & 0 & i \sqrt{5} & 0 & 0 \\
 0 & 0 & 0 & 0 & 0 & 0 & i & 0 & 0 & -i \sqrt{15} & 0 & 0 \\
 0 & 0 & 0 & 0 & 0 & 0 & 0 & 0 & 0 & 0 & 0 & 0 \\
 0 & 0 & 0 & 0 & 0 & 0 & 0 & 0 & 4 i & 0 & 0 & 0 \\
 0 & 0 & 0 & 0 & 0 & 0 & 0 & -4 i & 0 & 0 & 0 & 0 \\ \hline
 0 & 0 & 0 & 0 & 0 & 0 & 0 & 0 & 0 & 0 & 0 & 0 \\
 -i \sqrt{3} & -i & 0 & 0 & 0 & 0 & 0 & 0 & 0 & 0 & 0 & 0 \\
 0 & 0 & 0 & 0 & 4 i & 0 & 0 & 0 & 0 & 0 & 0 & 0 \\
 0 & 0 & 0 & -4 i & 0 & 0 & 0 & 0 & 0 & 0 & 0 & 0 \\
 -i \sqrt{5} & i \sqrt{15} & 0 & 0 & 0 & 0 & 0 & 0 & 0 & 0 & 0 & 0 \\
 0 & 0 & 0 & 0 & 0 & 0 & 0 & 0 & 0 & 0 & 0 & 0 \\
 0 & 0 & 0 & 0 & 0 & 0 & 0 & 0 & 0 & 0 & 0 & 0 \\
\end{array}
\right),
$}
\cr&
M_{zx}=\frac{2}{\sqrt{35}}
\scalebox{0.95}{$\displaystyle
\left(
\begin{array}{ccccc|ccccccc}
 0 & 0 & 0 & 0 & 0 & 0 & 0 & -i \sqrt{3} & 0 & 0 & i \sqrt{5} & 0 \\
 0 & 0 & 0 & 0 & 0 & 0 & 0 & i & 0 & 0 & i \sqrt{15} & 0 \\
 0 & 0 & 0 & 0 & 0 & 0 & 0 & 0 & -4 i & 0 & 0 & 0 \\
 0 & 0 & 0 & 0 & 0 & 0 & 0 & 0 & 0 & 0 & 0 & 0 \\
 0 & 0 & 0 & 0 & 0 & 0 & 4 i & 0 & 0 & 0 & 0 & 0 \\ \hline
 0 & 0 & 0 & 0 & 0 & 0 & 0 & 0 & 0 & 0 & 0 & 0 \\
 0 & 0 & 0 & 0 & -4 i & 0 & 0 & 0 & 0 & 0 & 0 & 0 \\
 i \sqrt{3} & -i & 0 & 0 & 0 & 0 & 0 & 0 & 0 & 0 & 0 & 0 \\
 0 & 0 & 4 i & 0 & 0 & 0 & 0 & 0 & 0 & 0 & 0 & 0 \\
 0 & 0 & 0 & 0 & 0 & 0 & 0 & 0 & 0 & 0 & 0 & 0 \\
 -i \sqrt{5} & -i \sqrt{15} & 0 & 0 & 0 & 0 & 0 & 0 & 0 & 0 & 0 & 0 \\
 0 & 0 & 0 & 0 & 0 & 0 & 0 & 0 & 0 & 0 & 0 & 0 \\
\end{array}
\right),
$}
\cr&
M_{xy}=\frac{4}{\sqrt{35}}
\scalebox{0.95}{$\displaystyle
\left(
\begin{array}{ccccc|ccccccc}
 0 & 0 & 0 & 0 & 0 & 0 & 0 & 0 & 0 & 0 & 0 & -i \sqrt{5} \\
 0 & 0 & 0 & 0 & 0 & 0 & 0 & 0 & -i & 0 & 0 & 0 \\
 0 & 0 & 0 & 0 & 0 & 0 & 0 & 2 i & 0 & 0 & 0 & 0 \\
 0 & 0 & 0 & 0 & 0 & 0 & -2 i & 0 & 0 & 0 & 0 & 0 \\
 0 & 0 & 0 & 0 & 0 & 0 & 0 & 0 & 0 & 0 & 0 & 0 \\ \hline
 0 & 0 & 0 & 0 & 0 & 0 & 0 & 0 & 0 & 0 & 0 & 0 \\
 0 & 0 & 0 & 2 i & 0 & 0 & 0 & 0 & 0 & 0 & 0 & 0 \\
 0 & 0 & -2 i & 0 & 0 & 0 & 0 & 0 & 0 & 0 & 0 & 0 \\
 0 & i & 0 & 0 & 0 & 0 & 0 & 0 & 0 & 0 & 0 & 0 \\
 0 & 0 & 0 & 0 & 0 & 0 & 0 & 0 & 0 & 0 & 0 & 0 \\
 0 & 0 & 0 & 0 & 0 & 0 & 0 & 0 & 0 & 0 & 0 & 0 \\
 i \sqrt{5} & 0 & 0 & 0 & 0 & 0 & 0 & 0 & 0 & 0 & 0 & 0 \\
\end{array}
\right),  
$}
\end{align}

\begin{align}
&
G_{u}=\frac{1}{\sqrt{35}}
\scalebox{0.95}{$\displaystyle
\left(
\begin{array}{ccccc|ccccccc}
 0 & 0 & 0 & 0 & 0 & 0 & 0 & 0 & 0 & 0 & 0 & 0 \\
 0 & 0 & 0 & 0 & 0 & 2 \sqrt{5} & 0 & 0 & 0 & 0 & 0 & 0 \\
 0 & 0 & 0 & 0 & 0 & 0 & \sqrt{3} & 0 & 0 & \sqrt{5} & 0 & 0 \\
 0 & 0 & 0 & 0 & 0 & 0 & 0 & -\sqrt{3} & 0 & 0 & \sqrt{5} & 0 \\
 0 & 0 & 0 & 0 & 0 & 0 & 0 & 0 & 0 & 0 & 0 & -2 \sqrt{5} \\ \hline
 0 & 2 \sqrt{5} & 0 & 0 & 0 & 0 & 0 & 0 & 0 & 0 & 0 & 0 \\
 0 & 0 & \sqrt{3} & 0 & 0 & 0 & 0 & 0 & 0 & 0 & 0 & 0 \\
 0 & 0 & 0 & -\sqrt{3} & 0 & 0 & 0 & 0 & 0 & 0 & 0 & 0 \\
 0 & 0 & 0 & 0 & 0 & 0 & 0 & 0 & 0 & 0 & 0 & 0 \\
 0 & 0 & \sqrt{5} & 0 & 0 & 0 & 0 & 0 & 0 & 0 & 0 & 0 \\
 0 & 0 & 0 & \sqrt{5} & 0 & 0 & 0 & 0 & 0 & 0 & 0 & 0 \\
 0 & 0 & 0 & 0 & -2 \sqrt{5} & 0 & 0 & 0 & 0 & 0 & 0 & 0 \\
\end{array}
\right), 
$}
\cr&
G_{v}=\frac{1}{\sqrt{35}}
\scalebox{0.95}{$\displaystyle
\left(
\begin{array}{ccccc|ccccccc}
 0 & 0 & 0 & 0 & 0 & -2 \sqrt{5} & 0 & 0 & 0 & 0 & 0 & 0 \\
 0 & 0 & 0 & 0 & 0 & 0 & 0 & 0 & 0 & 0 & 0 & 0 \\
 0 & 0 & 0 & 0 & 0 & 0 & 1 & 0 & 0 & -\sqrt{15} & 0 & 0 \\
 0 & 0 & 0 & 0 & 0 & 0 & 0 & 1 & 0 & 0 & \sqrt{15} & 0 \\
 0 & 0 & 0 & 0 & 0 & 0 & 0 & 0 & -2 & 0 & 0 & 0 \\ \hline
 -2 \sqrt{5} & 0 & 0 & 0 & 0 & 0 & 0 & 0 & 0 & 0 & 0 & 0 \\
 0 & 0 & 1 & 0 & 0 & 0 & 0 & 0 & 0 & 0 & 0 & 0 \\
 0 & 0 & 0 & 1 & 0 & 0 & 0 & 0 & 0 & 0 & 0 & 0 \\
 0 & 0 & 0 & 0 & -2 & 0 & 0 & 0 & 0 & 0 & 0 & 0 \\
 0 & 0 & -\sqrt{15} & 0 & 0 & 0 & 0 & 0 & 0 & 0 & 0 & 0 \\
 0 & 0 & 0 & \sqrt{15} & 0 & 0 & 0 & 0 & 0 & 0 & 0 & 0 \\
 0 & 0 & 0 & 0 & 0 & 0 & 0 & 0 & 0 & 0 & 0 & 0 \\
\end{array}
\right), 
$}
\cr&
G_{yz}=\frac{1}{\sqrt{35}}
\scalebox{0.95}{$\displaystyle
\left(
\begin{array}{ccccc|ccccccc}
 0 & 0 & 0 & 0 & 0 & 0 & -\sqrt{3} & 0 & 0 & -\sqrt{5} & 0 & 0 \\
 0 & 0 & 0 & 0 & 0 & 0 & -1 & 0 & 0 & \sqrt{15} & 0 & 0 \\
 0 & 0 & 0 & 0 & 0 & 0 & 0 & 0 & 0 & 0 & 0 & 0 \\
 0 & 0 & 0 & 0 & 0 & 0 & 0 & 0 & -4 & 0 & 0 & 0 \\
 0 & 0 & 0 & 0 & 0 & 0 & 0 & 4 & 0 & 0 & 0 & 0 \\ \hline
 0 & 0 & 0 & 0 & 0 & 0 & 0 & 0 & 0 & 0 & 0 & 0 \\
 -\sqrt{3} & -1 & 0 & 0 & 0 & 0 & 0 & 0 & 0 & 0 & 0 & 0 \\
 0 & 0 & 0 & 0 & 4 & 0 & 0 & 0 & 0 & 0 & 0 & 0 \\
 0 & 0 & 0 & -4 & 0 & 0 & 0 & 0 & 0 & 0 & 0 & 0 \\
 -\sqrt{5} & \sqrt{15} & 0 & 0 & 0 & 0 & 0 & 0 & 0 & 0 & 0 & 0 \\
 0 & 0 & 0 & 0 & 0 & 0 & 0 & 0 & 0 & 0 & 0 & 0 \\
 0 & 0 & 0 & 0 & 0 & 0 & 0 & 0 & 0 & 0 & 0 & 0 \\
\end{array}
\right),
$}
\cr&
G_{zx}=\frac{1}{\sqrt{35}}
\scalebox{0.95}{$\displaystyle
\left(
\begin{array}{ccccc|ccccccc}
 0 & 0 & 0 & 0 & 0 & 0 & 0 & \sqrt{3} & 0 & 0 & -\sqrt{5} & 0 \\
 0 & 0 & 0 & 0 & 0 & 0 & 0 & -1 & 0 & 0 & -\sqrt{15} & 0 \\
 0 & 0 & 0 & 0 & 0 & 0 & 0 & 0 & 4 & 0 & 0 & 0 \\
 0 & 0 & 0 & 0 & 0 & 0 & 0 & 0 & 0 & 0 & 0 & 0 \\
 0 & 0 & 0 & 0 & 0 & 0 & -4 & 0 & 0 & 0 & 0 & 0 \\ \hline
 0 & 0 & 0 & 0 & 0 & 0 & 0 & 0 & 0 & 0 & 0 & 0 \\
 0 & 0 & 0 & 0 & -4 & 0 & 0 & 0 & 0 & 0 & 0 & 0 \\
 \sqrt{3} & -1 & 0 & 0 & 0 & 0 & 0 & 0 & 0 & 0 & 0 & 0 \\
 0 & 0 & 4 & 0 & 0 & 0 & 0 & 0 & 0 & 0 & 0 & 0 \\
 0 & 0 & 0 & 0 & 0 & 0 & 0 & 0 & 0 & 0 & 0 & 0 \\
 -\sqrt{5} & -\sqrt{15} & 0 & 0 & 0 & 0 & 0 & 0 & 0 & 0 & 0 & 0 \\
 0 & 0 & 0 & 0 & 0 & 0 & 0 & 0 & 0 & 0 & 0 & 0 \\
\end{array}
\right),
$}
\cr&
G_{xy}=\frac{2}{\sqrt{35}}
\scalebox{0.95}{$\displaystyle
\left(
\begin{array}{ccccc|ccccccc}
 0 & 0 & 0 & 0 & 0 & 0 & 0 & 0 & 0 & 0 & 0 & \sqrt{5} \\
 0 & 0 & 0 & 0 & 0 & 0 & 0 & 0 & 1 & 0 & 0 & 0 \\
 0 & 0 & 0 & 0 & 0 & 0 & 0 & -2 & 0 & 0 & 0 & 0 \\
 0 & 0 & 0 & 0 & 0 & 0 & 2 & 0 & 0 & 0 & 0 & 0 \\
 0 & 0 & 0 & 0 & 0 & 0 & 0 & 0 & 0 & 0 & 0 & 0 \\ \hline
 0 & 0 & 0 & 0 & 0 & 0 & 0 & 0 & 0 & 0 & 0 & 0 \\
 0 & 0 & 0 & 2 & 0 & 0 & 0 & 0 & 0 & 0 & 0 & 0 \\
 0 & 0 & -2 & 0 & 0 & 0 & 0 & 0 & 0 & 0 & 0 & 0 \\
 0 & 1 & 0 & 0 & 0 & 0 & 0 & 0 & 0 & 0 & 0 & 0 \\
 0 & 0 & 0 & 0 & 0 & 0 & 0 & 0 & 0 & 0 & 0 & 0 \\
 0 & 0 & 0 & 0 & 0 & 0 & 0 & 0 & 0 & 0 & 0 & 0 \\
 \sqrt{5} & 0 & 0 & 0 & 0 & 0 & 0 & 0 & 0 & 0 & 0 & 0 \\
\end{array}
\right),  
$}
\end{align}

\begin{align}
&
Q_{xyz}=\frac{2}{3\sqrt{7}}
\left(
\begin{array}{ccccc|ccccccc}
 0 & 0 & 0 & 0 & 0 & 0 & 0 & 0 & 0 & 0 & 0 & 0 \\
 0 & 0 & 0 & 0 & 0 & 0 & 0 & 0 & 0 & 0 & 0 & 0 \\
 0 & 0 & 0 & 0 & 0 & 0 & -1 & 0 & 0 & 0 & 0 & 0 \\
 0 & 0 & 0 & 0 & 0 & 0 & 0 & -1 & 0 & 0 & 0 & 0 \\
 0 & 0 & 0 & 0 & 0 & 0 & 0 & 0 & -1 & 0 & 0 & 0 \\ \hline
 0 & 0 & 0 & 0 & 0 & 0 & 0 & 0 & 0 & 0 & 0 & 0 \\
 0 & 0 & -1 & 0 & 0 & 0 & 0 & 0 & 0 & 0 & 0 & 0 \\
 0 & 0 & 0 & -1 & 0 & 0 & 0 & 0 & 0 & 0 & 0 & 0 \\
 0 & 0 & 0 & 0 & -1 & 0 & 0 & 0 & 0 & 0 & 0 & 0 \\
 0 & 0 & 0 & 0 & 0 & 0 & 0 & 0 & 0 & 0 & 0 & 0 \\
 0 & 0 & 0 & 0 & 0 & 0 & 0 & 0 & 0 & 0 & 0 & 0 \\
 0 & 0 & 0 & 0 & 0 & 0 & 0 & 0 & 0 & 0 & 0 & 0 \\
\end{array}
\right), 
\cr&
Q_{x}^\alpha=\frac{1}{6\sqrt{35}}
\left(
\begin{array}{ccccc|ccccccc}
 0 & 0 & 0 & 0 & 0 & 0 & -4 & 0 & 0 & 2 \sqrt{15} & 0 & 0 \\
 0 & 0 & 0 & 0 & 0 & 0 & 4 \sqrt{3} & 0 & 0 & 2 \sqrt{5} & 0 & 0 \\
 0 & 0 & 0 & 0 & 0 & -4 \sqrt{5} & 0 & 0 & 0 & 0 & 0 & 0 \\
 0 & 0 & 0 & 0 & 0 & 0 & 0 & 0 & -\sqrt{3} & 0 & 0 & \sqrt{5} \\
 0 & 0 & 0 & 0 & 0 & 0 & 0 & -\sqrt{3} & 0 & 0 & -\sqrt{5} & 0 \\ \hline
 0 & 0 & -4 \sqrt{5} & 0 & 0 & 0 & 0 & 0 & 0 & 0 & 0 & 0 \\
 -4 & 4 \sqrt{3} & 0 & 0 & 0 & 0 & 0 & 0 & 0 & 0 & 0 & 0 \\
 0 & 0 & 0 & 0 & -\sqrt{3} & 0 & 0 & 0 & 0 & 0 & 0 & 0 \\
 0 & 0 & 0 & -\sqrt{3} & 0 & 0 & 0 & 0 & 0 & 0 & 0 & 0 \\
 2 \sqrt{15} & 2 \sqrt{5} & 0 & 0 & 0 & 0 & 0 & 0 & 0 & 0 & 0 & 0 \\
 0 & 0 & 0 & 0 & -\sqrt{5} & 0 & 0 & 0 & 0 & 0 & 0 & 0 \\
 0 & 0 & 0 & \sqrt{5} & 0 & 0 & 0 & 0 & 0 & 0 & 0 & 0 \\
\end{array}
\right), 
\cr&
Q_{y}^\alpha=\frac{1}{6\sqrt{35}}
\left(
\begin{array}{ccccc|ccccccc}
 0 & 0 & 0 & 0 & 0 & 0 & 0 & -4 & 0 & 0 & -2 \sqrt{15} & 0 \\
 0 & 0 & 0 & 0 & 0 & 0 & 0 & -4 \sqrt{3} & 0 & 0 & 2 \sqrt{5} & 0 \\
 0 & 0 & 0 & 0 & 0 & 0 & 0 & 0 & -\sqrt{3} & 0 & 0 & -\sqrt{5} \\
 0 & 0 & 0 & 0 & 0 & -4 \sqrt{5} & 0 & 0 & 0 & 0 & 0 & 0 \\
 0 & 0 & 0 & 0 & 0 & 0 & -\sqrt{3} & 0 & 0 & \sqrt{5} & 0 & 0 \\ \hline
 0 & 0 & 0 & -4 \sqrt{5} & 0 & 0 & 0 & 0 & 0 & 0 & 0 & 0 \\
 0 & 0 & 0 & 0 & -\sqrt{3} & 0 & 0 & 0 & 0 & 0 & 0 & 0 \\
 -4 & -4 \sqrt{3} & 0 & 0 & 0 & 0 & 0 & 0 & 0 & 0 & 0 & 0 \\
 0 & 0 & -\sqrt{3} & 0 & 0 & 0 & 0 & 0 & 0 & 0 & 0 & 0 \\
 0 & 0 & 0 & 0 & \sqrt{5} & 0 & 0 & 0 & 0 & 0 & 0 & 0 \\
 -2 \sqrt{15} & 2 \sqrt{5} & 0 & 0 & 0 & 0 & 0 & 0 & 0 & 0 & 0 & 0 \\
 0 & 0 & -\sqrt{5} & 0 & 0 & 0 & 0 & 0 & 0 & 0 & 0 & 0 \\
\end{array}
\right),
\cr&
Q_{z}^\alpha=\frac{1}{6\sqrt{35}}
\left(
\begin{array}{ccccc|ccccccc}
 0 & 0 & 0 & 0 & 0 & 0 & 0 & 0 & 8 & 0 & 0 & 0 \\
 0 & 0 & 0 & 0 & 0 & 0 & 0 & 0 & 0 & 0 & 0 & -4 \sqrt{5} \\
 0 & 0 & 0 & 0 & 0 & 0 & 0 & -\sqrt{3} & 0 & 0 & \sqrt{5} & 0 \\
 0 & 0 & 0 & 0 & 0 & 0 & -\sqrt{3} & 0 & 0 & -\sqrt{5} & 0 & 0 \\
 0 & 0 & 0 & 0 & 0 & -4 \sqrt{5} & 0 & 0 & 0 & 0 & 0 & 0 \\ \hline
 0 & 0 & 0 & 0 & -4 \sqrt{5} & 0 & 0 & 0 & 0 & 0 & 0 & 0 \\
 0 & 0 & 0 & -\sqrt{3} & 0 & 0 & 0 & 0 & 0 & 0 & 0 & 0 \\
 0 & 0 & -\sqrt{3} & 0 & 0 & 0 & 0 & 0 & 0 & 0 & 0 & 0 \\
 8 & 0 & 0 & 0 & 0 & 0 & 0 & 0 & 0 & 0 & 0 & 0 \\
 0 & 0 & 0 & -\sqrt{5} & 0 & 0 & 0 & 0 & 0 & 0 & 0 & 0 \\
 0 & 0 & \sqrt{5} & 0 & 0 & 0 & 0 & 0 & 0 & 0 & 0 & 0 \\
 0 & -4 \sqrt{5} & 0 & 0 & 0 & 0 & 0 & 0 & 0 & 0 & 0 & 0 \\
\end{array}
\right), \nonumber
\end{align}
\begin{align}
&
Q_{x}^\beta=\frac{1}{6\sqrt{35}}
\left(
\begin{array}{ccccc|ccccccc}
 0 & 0 & 0 & 0 & 0 & 0 & 2 \sqrt{15} & 0 & 0 & 0 & 0 & 0 \\
 0 & 0 & 0 & 0 & 0 & 0 & 2 \sqrt{5} & 0 & 0 & 0 & 0 & 0 \\
 0 & 0 & 0 & 0 & 0 & 0 & 0 & 0 & 0 & 0 & 0 & 0 \\
 0 & 0 & 0 & 0 & 0 & 0 & 0 & 0 & -\sqrt{5} & 0 & 0 & -5 \sqrt{3} \\
 0 & 0 & 0 & 0 & 0 & 0 & 0 & \sqrt{5} & 0 & 0 & -5 \sqrt{3} & 0 \\ \hline
 0 & 0 & 0 & 0 & 0 & 0 & 0 & 0 & 0 & 0 & 0 & 0 \\
 2 \sqrt{15} & 2 \sqrt{5} & 0 & 0 & 0 & 0 & 0 & 0 & 0 & 0 & 0 & 0 \\
 0 & 0 & 0 & 0 & \sqrt{5} & 0 & 0 & 0 & 0 & 0 & 0 & 0 \\
 0 & 0 & 0 & -\sqrt{5} & 0 & 0 & 0 & 0 & 0 & 0 & 0 & 0 \\
 0 & 0 & 0 & 0 & 0 & 0 & 0 & 0 & 0 & 0 & 0 & 0 \\
 0 & 0 & 0 & 0 & -5 \sqrt{3} & 0 & 0 & 0 & 0 & 0 & 0 & 0 \\
 0 & 0 & 0 & -5 \sqrt{3} & 0 & 0 & 0 & 0 & 0 & 0 & 0 & 0 \\
\end{array}
\right),
\cr&
Q_{y}^\beta=\frac{1}{6\sqrt{35}}
\left(
\begin{array}{ccccc|ccccccc}
 0 & 0 & 0 & 0 & 0 & 0 & 0 & -2 \sqrt{15} & 0 & 0 & 0 & 0 \\
 0 & 0 & 0 & 0 & 0 & 0 & 0 & 2 \sqrt{5} & 0 & 0 & 0 & 0 \\
 0 & 0 & 0 & 0 & 0 & 0 & 0 & 0 & \sqrt{5} & 0 & 0 & -5 \sqrt{3} \\
 0 & 0 & 0 & 0 & 0 & 0 & 0 & 0 & 0 & 0 & 0 & 0 \\
 0 & 0 & 0 & 0 & 0 & 0 & -\sqrt{5} & 0 & 0 & -5 \sqrt{3} & 0 & 0 \\ \hline
 0 & 0 & 0 & 0 & 0 & 0 & 0 & 0 & 0 & 0 & 0 & 0 \\
 0 & 0 & 0 & 0 & -\sqrt{5} & 0 & 0 & 0 & 0 & 0 & 0 & 0 \\
 -2 \sqrt{15} & 2 \sqrt{5} & 0 & 0 & 0 & 0 & 0 & 0 & 0 & 0 & 0 & 0 \\
 0 & 0 & \sqrt{5} & 0 & 0 & 0 & 0 & 0 & 0 & 0 & 0 & 0 \\
 0 & 0 & 0 & 0 & -5 \sqrt{3} & 0 & 0 & 0 & 0 & 0 & 0 & 0 \\
 0 & 0 & 0 & 0 & 0 & 0 & 0 & 0 & 0 & 0 & 0 & 0 \\
 0 & 0 & -5 \sqrt{3} & 0 & 0 & 0 & 0 & 0 & 0 & 0 & 0 & 0 \\
\end{array}
\right),
\cr&
Q_{z}^\beta=\frac{1}{6\sqrt{35}}
\left(
\begin{array}{ccccc|ccccccc}
 0 & 0 & 0 & 0 & 0 & 0 & 0 & 0 & 0 & 0 & 0 & 0 \\
 0 & 0 & 0 & 0 & 0 & 0 & 0 & 0 & -4 \sqrt{5} & 0 & 0 & 0 \\
 0 & 0 & 0 & 0 & 0 & 0 & 0 & -\sqrt{5} & 0 & 0 & -5 \sqrt{3} & 0 \\
 0 & 0 & 0 & 0 & 0 & 0 & \sqrt{5} & 0 & 0 & -5 \sqrt{3} & 0 & 0 \\
 0 & 0 & 0 & 0 & 0 & 0 & 0 & 0 & 0 & 0 & 0 & 0 \\ \hline
 0 & 0 & 0 & 0 & 0 & 0 & 0 & 0 & 0 & 0 & 0 & 0 \\
 0 & 0 & 0 & \sqrt{5} & 0 & 0 & 0 & 0 & 0 & 0 & 0 & 0 \\
 0 & 0 & -\sqrt{5} & 0 & 0 & 0 & 0 & 0 & 0 & 0 & 0 & 0 \\
 0 & -4 \sqrt{5} & 0 & 0 & 0 & 0 & 0 & 0 & 0 & 0 & 0 & 0 \\
 0 & 0 & 0 & -5 \sqrt{3} & 0 & 0 & 0 & 0 & 0 & 0 & 0 & 0 \\
 0 & 0 & -5 \sqrt{3} & 0 & 0 & 0 & 0 & 0 & 0 & 0 & 0 & 0 \\
 0 & 0 & 0 & 0 & 0 & 0 & 0 & 0 & 0 & 0 & 0 & 0 \\
\end{array}
\right),  
\end{align}

\begin{align}
&
T_{xyz}=\frac{1}{5\sqrt{7}}
\left(
\begin{array}{ccccc|ccccccc}
 0 & 0 & 0 & 0 & 0 & 0 & 0 & 0 & 0 & 0 & 0 & 0 \\
 0 & 0 & 0 & 0 & 0 & 0 & 0 & 0 & 0 & 0 & 0 & 0 \\
 0 & 0 & 0 & 0 & 0 & 0 & -i & 0 & 0 & 0 & 0 & 0 \\
 0 & 0 & 0 & 0 & 0 & 0 & 0 & -i & 0 & 0 & 0 & 0 \\
 0 & 0 & 0 & 0 & 0 & 0 & 0 & 0 & -i & 0 & 0 & 0 \\ \hline
 0 & 0 & 0 & 0 & 0 & 0 & 0 & 0 & 0 & 0 & 0 & 0 \\
 0 & 0 & i & 0 & 0 & 0 & 0 & 0 & 0 & 0 & 0 & 0 \\
 0 & 0 & 0 & i & 0 & 0 & 0 & 0 & 0 & 0 & 0 & 0 \\
 0 & 0 & 0 & 0 & i & 0 & 0 & 0 & 0 & 0 & 0 & 0 \\
 0 & 0 & 0 & 0 & 0 & 0 & 0 & 0 & 0 & 0 & 0 & 0 \\
 0 & 0 & 0 & 0 & 0 & 0 & 0 & 0 & 0 & 0 & 0 & 0 \\
 0 & 0 & 0 & 0 & 0 & 0 & 0 & 0 & 0 & 0 & 0 & 0 \\
\end{array}
\right), 
\cr&
T_{x}^\alpha=\frac{1}{20\sqrt{35}}
\left(
\begin{array}{ccccc|ccccccc}
 0 & 0 & 0 & 0 & 0 & 0 & -4 i & 0 & 0 & 2 i \sqrt{15} & 0 & 0 \\
 0 & 0 & 0 & 0 & 0 & 0 & 4 i \sqrt{3} & 0 & 0 & 2 i \sqrt{5} & 0 & 0 \\
 0 & 0 & 0 & 0 & 0 & -4 i \sqrt{5} & 0 & 0 & 0 & 0 & 0 & 0 \\
 0 & 0 & 0 & 0 & 0 & 0 & 0 & 0 & -i \sqrt{3} & 0 & 0 & i \sqrt{5} \\
 0 & 0 & 0 & 0 & 0 & 0 & 0 & -i \sqrt{3} & 0 & 0 & -i \sqrt{5} & 0 \\ \hline
 0 & 0 & 4 i \sqrt{5} & 0 & 0 & 0 & 0 & 0 & 0 & 0 & 0 & 0 \\
 4 i & -4 i \sqrt{3} & 0 & 0 & 0 & 0 & 0 & 0 & 0 & 0 & 0 & 0 \\
 0 & 0 & 0 & 0 & i \sqrt{3} & 0 & 0 & 0 & 0 & 0 & 0 & 0 \\
 0 & 0 & 0 & i \sqrt{3} & 0 & 0 & 0 & 0 & 0 & 0 & 0 & 0 \\
 -2 i \sqrt{15} & -2 i \sqrt{5} & 0 & 0 & 0 & 0 & 0 & 0 & 0 & 0 & 0 & 0 \\
 0 & 0 & 0 & 0 & i \sqrt{5} & 0 & 0 & 0 & 0 & 0 & 0 & 0 \\
 0 & 0 & 0 & -i \sqrt{5} & 0 & 0 & 0 & 0 & 0 & 0 & 0 & 0 \\
\end{array}
\right), 
\cr&
T_{y}^\alpha=\frac{1}{20\sqrt{35}}
\left(
\begin{array}{ccccc|ccccccc}
 0 & 0 & 0 & 0 & 0 & 0 & 0 & -4 i & 0 & 0 & -2 i \sqrt{15} & 0 \\
 0 & 0 & 0 & 0 & 0 & 0 & 0 & -4 i \sqrt{3} & 0 & 0 & 2 i \sqrt{5} & 0 \\
 0 & 0 & 0 & 0 & 0 & 0 & 0 & 0 & -i \sqrt{3} & 0 & 0 & -i \sqrt{5} \\
 0 & 0 & 0 & 0 & 0 & -4 i \sqrt{5} & 0 & 0 & 0 & 0 & 0 & 0 \\
 0 & 0 & 0 & 0 & 0 & 0 & -i \sqrt{3} & 0 & 0 & i \sqrt{5} & 0 & 0 \\ \hline
 0 & 0 & 0 & 4 i \sqrt{5} & 0 & 0 & 0 & 0 & 0 & 0 & 0 & 0 \\
 0 & 0 & 0 & 0 & i \sqrt{3} & 0 & 0 & 0 & 0 & 0 & 0 & 0 \\
 4 i & 4 i \sqrt{3} & 0 & 0 & 0 & 0 & 0 & 0 & 0 & 0 & 0 & 0 \\
 0 & 0 & i \sqrt{3} & 0 & 0 & 0 & 0 & 0 & 0 & 0 & 0 & 0 \\
 0 & 0 & 0 & 0 & -i \sqrt{5} & 0 & 0 & 0 & 0 & 0 & 0 & 0 \\
 2 i \sqrt{15} & -2 i \sqrt{5} & 0 & 0 & 0 & 0 & 0 & 0 & 0 & 0 & 0 & 0 \\
 0 & 0 & i \sqrt{5} & 0 & 0 & 0 & 0 & 0 & 0 & 0 & 0 & 0 \\
\end{array}
\right),
\cr&
T_{z}^\alpha=\frac{1}{20\sqrt{35}}
\left(
\begin{array}{ccccc|ccccccc}
 0 & 0 & 0 & 0 & 0 & 0 & 0 & 0 & 8 i & 0 & 0 & 0 \\
 0 & 0 & 0 & 0 & 0 & 0 & 0 & 0 & 0 & 0 & 0 & -4 i \sqrt{5} \\
 0 & 0 & 0 & 0 & 0 & 0 & 0 & -i \sqrt{3} & 0 & 0 & i \sqrt{5} & 0 \\
 0 & 0 & 0 & 0 & 0 & 0 & -i \sqrt{3} & 0 & 0 & -i \sqrt{5} & 0 & 0 \\
 0 & 0 & 0 & 0 & 0 & -4 i \sqrt{5} & 0 & 0 & 0 & 0 & 0 & 0 \\ \hline
 0 & 0 & 0 & 0 & 4 i \sqrt{5} & 0 & 0 & 0 & 0 & 0 & 0 & 0 \\
 0 & 0 & 0 & i \sqrt{3} & 0 & 0 & 0 & 0 & 0 & 0 & 0 & 0 \\
 0 & 0 & i \sqrt{3} & 0 & 0 & 0 & 0 & 0 & 0 & 0 & 0 & 0 \\
 -8 i & 0 & 0 & 0 & 0 & 0 & 0 & 0 & 0 & 0 & 0 & 0 \\
 0 & 0 & 0 & i \sqrt{5} & 0 & 0 & 0 & 0 & 0 & 0 & 0 & 0 \\
 0 & 0 & -i \sqrt{5} & 0 & 0 & 0 & 0 & 0 & 0 & 0 & 0 & 0 \\
 0 & 4 i \sqrt{5} & 0 & 0 & 0 & 0 & 0 & 0 & 0 & 0 & 0 & 0 \\
\end{array}
\right), \nonumber
\end{align}
\begin{align}
&
T_{x}^\beta=\frac{1}{20\sqrt{35}}
\left(
\begin{array}{ccccc|ccccccc}
 0 & 0 & 0 & 0 & 0 & 0 & 2 i \sqrt{15} & 0 & 0 & 0 & 0 & 0 \\
 0 & 0 & 0 & 0 & 0 & 0 & 2 i \sqrt{5} & 0 & 0 & 0 & 0 & 0 \\
 0 & 0 & 0 & 0 & 0 & 0 & 0 & 0 & 0 & 0 & 0 & 0 \\
 0 & 0 & 0 & 0 & 0 & 0 & 0 & 0 & -i \sqrt{5} & 0 & 0 & -5 i \sqrt{3} \\
 0 & 0 & 0 & 0 & 0 & 0 & 0 & i \sqrt{5} & 0 & 0 & -5 i \sqrt{3} & 0 \\ \hline
 0 & 0 & 0 & 0 & 0 & 0 & 0 & 0 & 0 & 0 & 0 & 0 \\
 -2 i \sqrt{15} & -2 i \sqrt{5} & 0 & 0 & 0 & 0 & 0 & 0 & 0 & 0 & 0 & 0 \\
 0 & 0 & 0 & 0 & -i \sqrt{5} & 0 & 0 & 0 & 0 & 0 & 0 & 0 \\
 0 & 0 & 0 & i \sqrt{5} & 0 & 0 & 0 & 0 & 0 & 0 & 0 & 0 \\
 0 & 0 & 0 & 0 & 0 & 0 & 0 & 0 & 0 & 0 & 0 & 0 \\
 0 & 0 & 0 & 0 & 5 i \sqrt{3} & 0 & 0 & 0 & 0 & 0 & 0 & 0 \\
 0 & 0 & 0 & 5 i \sqrt{3} & 0 & 0 & 0 & 0 & 0 & 0 & 0 & 0 \\
\end{array}
\right),
\cr&
T_{y}^\beta=\frac{1}{20\sqrt{35}}
\left(
\begin{array}{ccccc|ccccccc}
 0 & 0 & 0 & 0 & 0 & 0 & 0 & -2 i \sqrt{15} & 0 & 0 & 0 & 0 \\
 0 & 0 & 0 & 0 & 0 & 0 & 0 & 2 i \sqrt{5} & 0 & 0 & 0 & 0 \\
 0 & 0 & 0 & 0 & 0 & 0 & 0 & 0 & i \sqrt{5} & 0 & 0 & -5 i \sqrt{3} \\
 0 & 0 & 0 & 0 & 0 & 0 & 0 & 0 & 0 & 0 & 0 & 0 \\
 0 & 0 & 0 & 0 & 0 & 0 & -i \sqrt{5} & 0 & 0 & -5 i \sqrt{3} & 0 & 0 \\ \hline
 0 & 0 & 0 & 0 & 0 & 0 & 0 & 0 & 0 & 0 & 0 & 0 \\
 0 & 0 & 0 & 0 & i \sqrt{5} & 0 & 0 & 0 & 0 & 0 & 0 & 0 \\
 2 i \sqrt{15} & -2 i \sqrt{5} & 0 & 0 & 0 & 0 & 0 & 0 & 0 & 0 & 0 & 0 \\
 0 & 0 & -i \sqrt{5} & 0 & 0 & 0 & 0 & 0 & 0 & 0 & 0 & 0 \\
 0 & 0 & 0 & 0 & 5 i \sqrt{3} & 0 & 0 & 0 & 0 & 0 & 0 & 0 \\
 0 & 0 & 0 & 0 & 0 & 0 & 0 & 0 & 0 & 0 & 0 & 0 \\
 0 & 0 & 5 i \sqrt{3} & 0 & 0 & 0 & 0 & 0 & 0 & 0 & 0 & 0 \\
\end{array}
\right),
\cr&
T_{z}^\beta=\frac{1}{20\sqrt{35}}
\left(
\begin{array}{ccccc|ccccccc}
 0 & 0 & 0 & 0 & 0 & 0 & 0 & 0 & 0 & 0 & 0 & 0 \\
 0 & 0 & 0 & 0 & 0 & 0 & 0 & 0 & -4 i \sqrt{5} & 0 & 0 & 0 \\
 0 & 0 & 0 & 0 & 0 & 0 & 0 & -i \sqrt{5} & 0 & 0 & -5 i \sqrt{3} & 0 \\
 0 & 0 & 0 & 0 & 0 & 0 & i \sqrt{5} & 0 & 0 & -5 i \sqrt{3} & 0 & 0 \\
 0 & 0 & 0 & 0 & 0 & 0 & 0 & 0 & 0 & 0 & 0 & 0 \\ \hline
 0 & 0 & 0 & 0 & 0 & 0 & 0 & 0 & 0 & 0 & 0 & 0 \\
 0 & 0 & 0 & -i \sqrt{5} & 0 & 0 & 0 & 0 & 0 & 0 & 0 & 0 \\
 0 & 0 & i \sqrt{5} & 0 & 0 & 0 & 0 & 0 & 0 & 0 & 0 & 0 \\
 0 & 4 i \sqrt{5} & 0 & 0 & 0 & 0 & 0 & 0 & 0 & 0 & 0 & 0 \\
 0 & 0 & 0 & 5 i \sqrt{3} & 0 & 0 & 0 & 0 & 0 & 0 & 0 & 0 \\
 0 & 0 & 5 i \sqrt{3} & 0 & 0 & 0 & 0 & 0 & 0 & 0 & 0 & 0 \\
 0 & 0 & 0 & 0 & 0 & 0 & 0 & 0 & 0 & 0 & 0 & 0 \\
\end{array}
\right). 
\end{align}

\subsection{$s$-$d$ hybridized-orbital system}
Matrix elements of each multipole activated in the $s$-$d$ hybridized-orbital system are given below. 
The basis function is taken as $(\phi_0,\phi_{u},\phi_{v},\phi_{yz},\phi_{zx},\phi_{xy})$.

\begin{align}
&
M_{x}=
\left(
\begin{array}{c|ccccc}
 0 & 0 & 0 & 0 & 0 & 0 \\ \hline
 0 & 0 & 0 & i \sqrt{3} & 0 & 0 \\
 0 & 0 & 0 & i & 0 & 0 \\
 0 & -i \sqrt{3} & -i & 0 & 0 & 0 \\
 0 & 0 & 0 & 0 & 0 & i \\
 0 & 0 & 0 & 0 & -i & 0 \\
\end{array}
\right),
M_{y}=
\left(
\begin{array}{c|ccccc}
 0 & 0 & 0 & 0 & 0 & 0 \\ \hline
 0 & 0 & 0 & 0 & -i \sqrt{3} & 0 \\
 0 & 0 & 0 & 0 & i & 0 \\
 0 & 0 & 0 & 0 & 0 & -i \\
 0 & i \sqrt{3} & -i & 0 & 0 & 0 \\
 0 & 0 & 0 & i & 0 & 0 \\
\end{array}
\right),
M_{z}=
\left(
\begin{array}{c|ccccc}
 0 & 0 & 0 & 0 & 0 & 0 \\ \hline
 0 & 0 & 0 & 0 & 0 & 0 \\
 0 & 0 & 0 & 0 & 0 & -2 i \\
 0 & 0 & 0 & 0 & i & 0 \\
 0 & 0 & 0 & -i & 0 & 0 \\
 0 & 0 & 2 i & 0 & 0 & 0 \\
\end{array}%
\right).
\end{align}

\subsection{$p$-$f$ hybridized-orbital system}
Matrix elements of each multipole activated in the $p$-$f$ hybridized-orbital system are given below. 
The basis function is taken as $(\phi_{x},\phi_{y},\phi_{z},\phi_{xyz},\phi_{x}^{\alpha},\phi_{y}^{\alpha},\phi_{z}^{\alpha},\phi_{x}^{\beta},\phi_{y}^{\beta},\phi_{z}^{\beta})$.

\begin{align}
&
M_{x}=\frac{1}{2}
\left(
%
\right).
\end{align}

\end{widetext}
\end{document}